\documentclass[notitlepage,showpacs,amsmath,amssymb,groupedaddress,longbibliography]{revtex4-2}
\usepackage{latexsym,amsmath,amssymb}
\usepackage{verbatim}
\usepackage{siunitx}
\usepackage[bookmarks=false,linkcolor=blue,urlcolor=blue,colorlinks,citecolor=blue]{hyperref}

\usepackage{graphicx}
\usepackage[dvipsnames]{xcolor}
\usepackage{color}
\usepackage{xspace}
\usepackage{soul}

\usepackage{dcolumn}
\usepackage{bm}

\renewcommand{\selectlanguage}[1]{}

\setcitestyle{super,open={},close={}}

\newcommand{\vbf}[1]{{\mathbf{ #1 }}}
\newcommand{\absval}[1]{{\left\vert #1 \right\vert}}

\newcommand{\momA}       {\left( \nabla - i \frac{2e}{\hbar c} \mathbf{A} \right)}

\begin{document}
\newcommand{\ovec}{\overrightarrow}

\title{Perfect supercurrent diode efficiency in chiral nanotube-based weak links}
\author{Joseph~J.~Cuozzo}
\email{jjcuozzo@utep.edu}
\affiliation{Department of Physics, The University of Texas at El Paso, El Paso, TX, USA.}\affiliation{Materials Physics Department, Sandia National Laboratories, Livermore, CA 94551, USA.}
\author{Fran\c cois L\'{e}onard}
\affiliation{Materials Physics Department, Sandia National Laboratories, Livermore, CA 94551, USA.}

\maketitle 

{\bf
     The supercurrent diode effect (SDE) describes superconducting systems where the magnitude of the superconducting-to-normal state switching current differs for positive and negative current bias.
     Despite the ubiquity of such diode effects in Josephson devices, the fundamental conditions to observe a diode effect in a Josephson junction and achieve perfect diode efficiency remain unclear.
     In this work, we analyze the supercurrent diode properties of a chiral nanotube-based weak link within a Ginzburg-Landau theory. 
     We find a diode effect and anomalous phase develop across the junction when a magnetic field is applied parallel to the tube despite the absence of spin-orbit interactions in the system. Unexpectedly, the SDE in the junction is independent of the anomalous phase. Alternatively, we determine a non-reciprocal persistent current that is protected by fluxoid quantization can activate SDE, even in the absence of higher-order pair tunneling processes. We show this new type of SDE can lead to, in principle, a perfect diode efficiency, highlighting how persistent currents can be used to engineer high efficiency supercurrent diodes.
}\\

\section{Introduction}
\begin{figure}[b]
\centering
\includegraphics[width=0.6\linewidth]{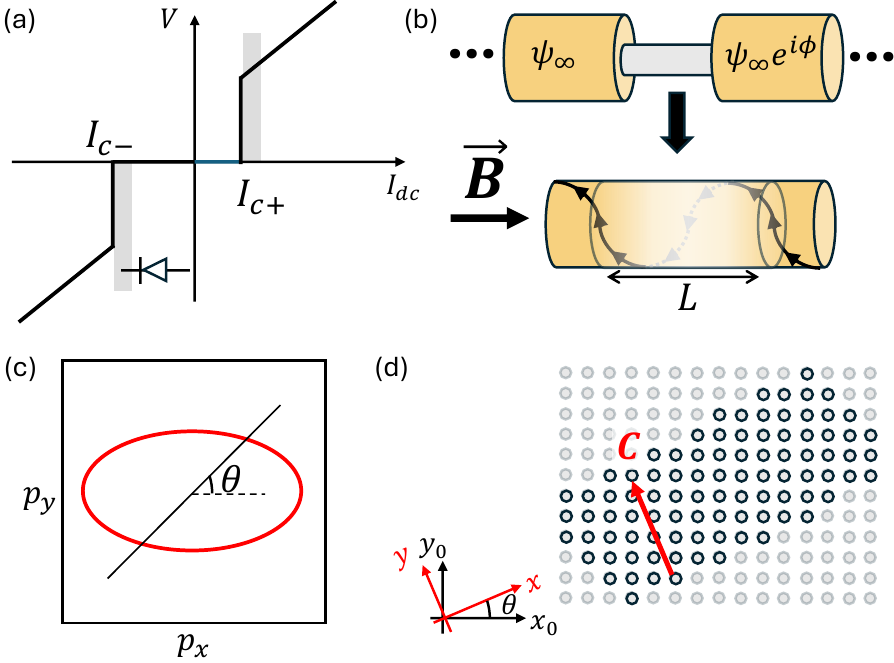}
\caption{\label{fig:setup}{\bf Chiral nanotube-based supercurrent diode:}
(a) Schematic of voltage-current curve of a supercurrent diode having a negative diode polarity.
(b) Cartoon of the ChNt-JJs and the helical persistent current inducing the Josephson diode effect.
(c) Schematic of $\mathcal{C}_2$-symmetric Fermi surface.
(d) Illustration of chiral nanotube on a square lattice.
}
\end{figure}

With a growing need for low-power fast electronics with low operational temperatures, a rapid development of non-reciprocal superconducting devices has occurred recently~\cite{Zhang2022, Nadeem2023}.
Non-reciprocal effects have been intensely investigated in bulk superconductors~\cite{Hu2007, Halterman2022, Wu2022, narita_field-free_2022, Hou2023, He2022, He2023} as well as in Josephson junctions~\cite{Shi2015, Bocquillon2017, Pal2019, Misaki2021, Baumgartner2022_JPhys, Baumgartner2022_NatNano, jeon_zero-field_2022, Pal2022, Kokkeler2022, Davydova2022, Illic2022_PRApplied, Tanaka2022, Trahms2023, Cayao2024, Yerin2024, Lu2023, Debnath2024, Debnath2025} (JJs)-- two superconducting electrodes weakly coupled by a tunneling barrier, such as a normal metal or insulator.
In JJs, the Josephson diode effect (JDE) manifests as a difference in magnitude of positive and negative threshold currents where the device switches from a superconducting to a dissipative state, shown schematically in Fig.~\ref{fig:setup}(a). Despite the effect having been known decades ago in devices with geometric inhomogeneities, the subject has received renewed interest based on superconducting diode effects reported in homogeneous devices where non-reciprocity arises from microscopic interactions. In these systems the diode effect, particularly when no magnetic field is applied, has supported the discovery of exotic states, such as time-reversal symmetry broken superconducting states~\cite{Zhao2023, Yu2024}. Other zero-field supercurrent diodes include JJs made of iron-based superconductors \cite{qiu_emergent_2023}, JJs of twisted bilayer graphene \cite{diez-merida_symmetry-broken_2023}, twisted trilayer graphene \cite{lin_zero-field_2022}, obstructed atomic insulator JJs \cite{wu_field-free_2022}, strained PbTaSe$_2$~\cite{Liu2024_strained}, and multiferroic JJs~\cite{Yang2025}.
Open questions linger about the nature of the SDE in some of these systems and others, which calls for further theoretical modeling to address the fundamental limits of the SDE.

Theoretical descriptions of the JDE often heuristically focus on symmetry arguments, namely, broken inversion and time-reversal symmetries~\cite{Onsager1931, Kubo1957, Rikken2001, Zhang2022}. Assuming these symmetries are broken, one may consider a common minimal expression for the current-phase relationship (CPR)~\cite{Nadeem2023}:
$
I_s(\phi) = a \sin(\phi) + b \cos(\phi) + c \sin(2\phi) + d\cos(2\phi).
$
Here $\phi$ is the phase across the Josephson junction, and $a,b,c$ and $d$ are real-valued constants describing the weights describing phase-coherent Cooper pair tunneling ($a$ and $b$) and pair co-tunneling ($c$ and $d$) supercurrent channels. The parameters $b$ and $d$ are associated with broken time-reversal symmetry (TRS) in the junction.
When all four constants are non-zero and treated as independent parameters, then ubiquitous combinations of $(a,b,c,d)$ result in $I_{c+} \ne \vert I_{c-} \vert$, where
$
I_{c+} = \max\left(I_s \right)
$
and 
$
I_{c-} = \min\left(I_s \right),
$
and the Josephson diode effect is realized (see Fig.~\ref{fig:setup}(a)) with an efficiency $\eta = \frac{I_{c+} + I_{c-}}{I_{c+} - I_{c-}}$.
Using the minimal form of $I_s(\phi)$, it is apparently not possible to have a perfectly efficient supercurrent diode where either $I_{c+}$ or $I_{c-}$ are zero. While ideal diode operation has been identified in the ac limit~\cite{Cuozzo2024_squid, Valentini2024, Souto2024}, there remains a question in the dc case: is it possible, in principle, to have a perfectly efficient Josephson diode operating in the dc limit? We answer this question in the affirmative by considering chiral nanotube-based weak link (ChNt-WLs) which serves as a model for short quasi-1D JJs.

In this work, we present a phenomenological Ginzburg-Landau (GL) theory for a ChNt-WL with a magnetic field applied along the tube, see Fig.~\ref{fig:setup}(b). 
We consider an anisotropic free energy functional that obeys inversion symmetry and choose periodic boundary conditions along non-high symmetry axes to define a chiral nanotube, see Fig.~\ref{fig:setup}(c-d). We first show by numerically solving the GL equations that a ChNt-WL can have a near-perfect SDE. We then analyze the system in the short radius limit where the CPR can be treated analytically.
We show that the ChNt-WL develops an anomalous phase despite the absence of spin-orbit coupling. 
The link between the anomalous phase and JDE has been investigated in JJs that break inversion symmetry~\cite{Reinhardt2024}, but the existence and possible origin of an anomalous phase in ChNt-WLs has not been identified until now. 
We also find a diode effect in the absence of pair co-tunneling, due to a non-reciprocal persistent current which is protected by fluxoid quantization in the ChNt-WL. In this case, a phase-independent persistent current allows us to place an upper bound on the diode efficiency demonstrating how perfect efficiency can be achieved in ChNt-WLs.

\section{Model}
We model a superconducting ChNt with higher-order terms in the free energy functional~\cite{He2023}:
\begin{align}
    F[\psi] - F[0] &= \int_{\Omega} \mathbf{dr}_0 \left(\alpha \vert \psi \vert^2 + \frac{\beta}{2} \vert \psi\vert^4 + \frac{1}{2m_0}\vert \mathbf{p}_0 \psi \vert^2 + \frac{1}{4m_0^2\zeta_0} \absval{\vbf{p}_0^2 \psi}^2   \right) + \int_{\Omega} \mathbf{dr}_0  \frac{1}{2m_1}\left( \vert p_{x0} \psi \vert^2 - \vert p_{y0} \psi \vert^2 \right) \nonumber \\
    &+ \int_{\Omega} \mathbf{dr}_0  \frac{\absval{p_{x0}^2 \psi}^2 + \absval{p_{y0}^2\psi}^2 - \frac{1}{2} \absval{\{p_{x0},p_{y0}\} \psi}^2}{4 m_1^2 \zeta_1}+ \int_{\Omega} \mathbf{dr}_0 \frac{\absval{p_{x0}^2 \psi}^2 + \absval{p_{y0}^2\psi}^2 - \frac{3}{2} \absval{\{p_{x0}, p_{y0}\} \psi}^2}{4 m_2^2 \zeta_2}.
    \label{eq:nt_F}
\end{align}
Here  $m_0 < m_1,m_2$, $\zeta_0, \zeta_1, \zeta_2 > 0$, $\alpha \propto (T - T_c)$ and $\beta$ are the usual GL coefficients, and
$
\mathbf{p} = -i\hbar \momA
$
is the momentum operator.
The atomic lattice of the system is defined on a 2D sheet $\Omega \in \mathbb{R}^2$. 
Here the kinetic energy contribution to $F$ has inversion symmetry and $\mathcal{C}_2$ rotational symmetry when 
$
1/m_1 \ne 0.
$
This corresponds to an elliptical Fermi surface, see Fig.~\ref{fig:setup}(c).
The circumferential vector 
$
\mathbf{C} = 2\pi R (-\sin\theta,\cos\theta),
$
where $R$ is the radius of the nanotube, defines the periodic boundary conditions to apply to the nanotube, see Fig.~\ref{fig:setup}(d). To simplify our analysis, we rotate our real space basis 
$
\mathbf{r} = R(\theta) \mathbf{r}_0
$ by the chiral angle $\theta$ where $R(\theta)$ is the 2x2 rotation matrix.
The Ginzburg-Landau equations in this case are (see Supplementary Information for details)
\begin{align}
    & \mathbf{J}  = \frac{2e\hbar }{i} 
    \begin{pmatrix}
        \rho_1 -\rho_4 p_x^2 - \rho_7 p_y^2 - \rho_8 p_x p_y && \rho_3 -\frac{\rho_8}{2} ( p_x^2 - p_y^2 )- \frac{\rho_6}{2} p_x p_y) \\
        \rho_3 -\frac{\rho_8}{2} ( p_x^2 - p_y^2 )- \frac{\rho_6}{2} p_x p_y) && \rho_2 - \rho_4 p_y^2 - \rho_7 p_x^2 + \rho_8  p_x p_y
    \end{pmatrix} \mathbf{j} \label{eq:Jnt_2} \\
    & \left[\alpha + \beta \absval{\psi}^2 - \left( \rho_1 p_x^2 + \rho_2 p_y^2 + 2 \rho_3 p_x p_y \right)  +  \rho_4 \left(p_x^4 + p_y^4 \right) + \left(\rho_6 + 2 \rho_7\right) p_x^2 p_y^2 + 2 \rho_8 \left(p_x^3 p_y - p_x p_y^3\right) \right]\psi = 0, \label{eq:GLnt_2}
\end{align}
where
$
\mathbf{j} = \psi^* \nabla \psi - \psi \nabla \psi^* - i \frac{4e}{\hbar c} \mathbf{A} \vert \psi\vert^2.
$ 
The coefficients associated with the reduced $\mathcal{C}_2$ symmetry are
$
\rho_1 = \mu_1 \cos^2\theta + \mu_2 \sin^2\theta,
$
$
\rho_2 = \mu_2 \cos^2\theta + \mu_1 \sin^2\theta,
$
and 
$
\rho_3 = (\mu_2 - \mu_1)  \sin 2\theta
$
where 
$
\mu_{1/2} = (m_1 \pm m_0)/(2m_0 m_1).
$
Higher order kinetic energy terms in the free energy determine the coefficients
$\rho_4 = (\kappa_1 + \lambda \cos 4\theta)/2$,
$\rho_6 = -2(\kappa_2 + \lambda \cos 4\theta)$,
$\rho_7 = (\kappa_1 + 2 \kappa_2 - \lambda \cos 4\theta)/2$,
and $\rho_8 = -\lambda \sin 4\theta$ where
$\kappa_1 = \frac{1}{2m_0^2 \zeta_0}+ \frac{1}{4 m_1^2 \zeta_1}$,
$\kappa_2 = \frac{1}{4m_2^2 \zeta_2}$, 
and $\lambda = \frac{1}{4m_1^2 \zeta_1}+ \frac{1}{2 m_2^2 \zeta_2}$.
When $\theta \mod \frac{\pi}{2} \ne 0$, the nanotube is chiral and $\rho_3 \ne 0$, causing broken mirror symmetry along the nanotube. 
Here the superfluid stiffness tensor, relating the supercurrent density $\bf J$ to the condensate current $\bf j$, now has a $\bf p$-dependence. 

To establish the conditions for the SDE, we consider an external magnetic field along the nanotube $\mathbf{B} = B_{ext}\mathbf{x}$ which breaks time-reversal symmetry. Periodic boundary conditions lead to $p_y$ being a good quantum number 
$
p_{y}^{(n)} = \hbar (\pi R^2 B_{ext}/\Phi_0 + n)/R = \hbar (\hat{\Phi}+n)/R
$
for some $n \in \mathbb{Z}$ and where $\Phi_0$ is the flux quantum and $\hat{\Phi} = \Phi/\Phi_0$. 
From here, to calculate the supercurrent in a Josephson junction, we need to apply appropriate boundary conditions to the solution to Eq.~(\ref{eq:GLnt_2}) in the weak link.
The simplest model for a narrow constriction uses rigid boundary conditions~\cite{Buzdin2008}
\begin{align}
    \psi(x\le 0) = \psi_{\infty}, \quad \psi(x \ge L) = \psi_{\infty} e^{i\phi},
\end{align}
where the region defining the junction is $0 \le x \le L$, $\phi$ is the phase difference across the junction, and 
$
\psi_{\infty} = \sqrt{-\alpha / \beta}.
$
The supercurrent density calculated in the weak link region describes the flow of Cooper pairs, both with and without $\phi$-sensitivity. The $\phi$-dependent contribution to the supercurrent density describes phase-coherent transport of Cooper pairs across the junction, which is the Josephson current. The $\phi$-independent term is a persistent current. This approach within GL theory qualitatively captures the current-phase relationship of Josephson junction~\cite{Likharev1979} and has been successfully used before to describe Rashba JJs under an applied magnetic field~{\cite{Buzdin2008}}.
A more rigorous treatment of the boundary conditions with a self-consistent treatment of $\psi$ for the junction can be made and may be important in 2D and 3D geometries~\cite{Kochan2023}, but for our quasi-1D system rigid boundary conditions provide sufficient qualitative accuracy~\cite{Likharev1979}.
In superconductor-quantum dot-superconductor JJs, Coulomb interactions have been predicted to influence the CPR and induce a zero field SDE~{\cite{Debnath2024, Debnath2025}}. In our model, we assume the Josephson energy of the weak link is greater than the charging energy so that the phase difference across the junction can be treated classically and effects of Coulomb interactions are negligible.
Here we also assume $T \lesssim T_c$, and take the short junction limit $L \ll \xi$ where 
$\xi = \sqrt{\hbar^2/(2m_0\vert \alpha \vert)}$
is the GL superconducting coherence length.

\begin{figure}[t]
\centering
\includegraphics[width=0.6\linewidth]{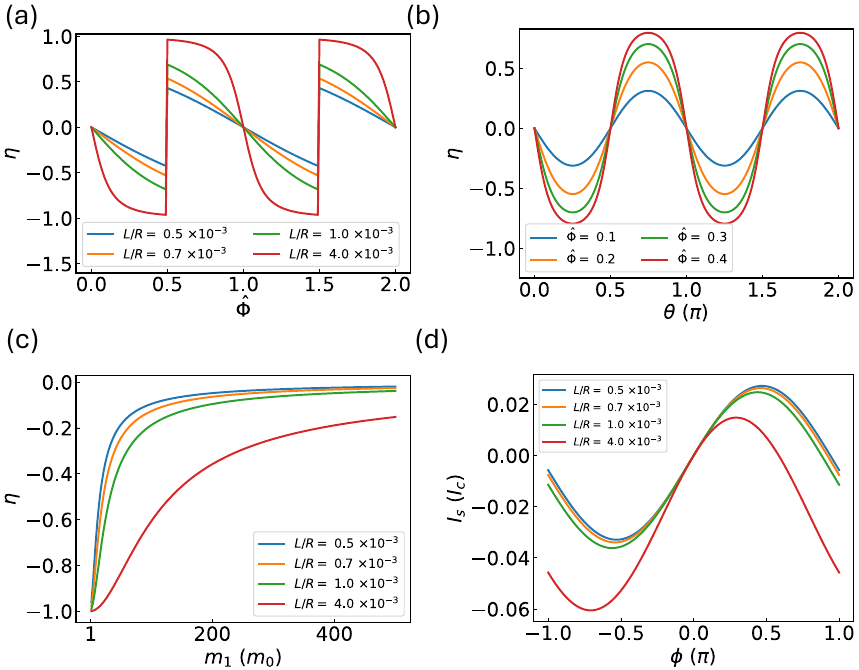}
\caption{\label{fig:fig2}{\bf Approaching perfect supercurrent diode efficiency:}
(a) $\eta$ versus $\hat{\Phi}$ for various ratios of $L/R$, $m_1/m_0 = 100$, and $\theta = 0.1\pi$.
(b) $\eta$ versus $\theta$ for $\hat{\Phi} = 0.1,~0.2,~0.3,~0.4$, $m_1/m_0 = 100$, and $L/R=10^{-3}$.
(c) $\eta$ versus $m_1$ for various ratios of $L/R$ and with $\hat{\Phi} = 0.1$ and $\theta = 0.1\pi$.
(d) CPR of the junction for various ratios of $L/R$ with $\hat{\Phi} = 0.1$, $m_1/m_0 = 100$, and $\theta = 0.1\pi$ using the normalization $I_c = \frac{4e\hbar \rho_1 \psi_{\infty}^2 A_{\perp}}{L}$.
We numerically solved Eq.~(\ref{eq:Jnt_2}-\ref{eq:GLnt_2}) as described in the main text with $\frac{\hbar^2}{m_0 \zeta_0 L^2} = 2 \times 10^{-3}$, $\frac{\hbar^2 m_0}{m_1^2 \zeta_0 L^2} = \frac{\hbar^2 m_0}{m_1^2 \zeta_0 L^2} = 2 \times 10^{-4}$, and $\frac{\alpha m_0 L^2}{\hbar^2} = 10^{-4}$. 
}
\end{figure}

\section{Results}

\subsection{Numerical solution}
We solve Eq.~(\ref{eq:GLnt_2}) with separation of variables using the order parameter
\begin{align}
    \psi_n(x,y) & = \psi_x(x) e^{ip_y^{(n)} y/\hbar},
\end{align}
for some complex-valued function $\psi_x$. In the short junction limit we can ignore the cubic $\vert \psi \vert^2 \psi$ term in Eq.~(\ref{eq:GLnt_2}) and reduce the equation to a linear fourth-order differential equation for $\psi_x$. In the following, we will also ignore $d^3\psi_x/dx^3$ and $d^4\psi_x/dx^4$ terms in solving the differential equation for two reasons: the coefficients of these terms are negligibly small in all simulations we present, and $d^m\psi_x/dx^m$ for $m>2$ are small in the short junction limit. We have numerically verified the numerical solutions we find are consistent with these approximations. In order to calculate the diode efficiency, we numerically calculate $\psi_x(x)$ to find $\psi_n(x,y)$ and then, following the application of boundary conditions, we calculate the supercurrent density $J_x(\phi,n)$ using Eq.~(\ref{eq:Jnt_2}). Integrating $J_x(\phi,n)$ over the junction region, we calculate the CPR $I_s(\phi,n_{eq})$ by minimizing the free energy in Eq.~(\ref{eq:nt_F}) with respect to $n$ and $\phi$ to find the value of $n$ at equilibrium, $n_{eq}$, and then treat $\phi$ as an externally tunable parameter.

In Fig.~\ref{fig:fig2}(a), we present the diode efficiency $\eta$ of the ChNt-WL as a function of the normalized magnetic flux threading the nanotube $\hat{\Phi}$. Here we take $\theta = 0.1\pi$ and the radius of the nanotube $R$ to be large compared to the length of the junction $L$. We find the magnetic field dependence of $\eta$ exhibits a complete suppression of $\eta$ when the magnetic flux in the nanotube is a half-integer multiple of $\Phi_0$, and $\Phi_0$-periodic oscillations. The periodicity of $\eta(\hat{\Phi})$ is a consequence of the Little-Parks effect. We also find $\eta$ is weakened with increasing $R$, but remarkably we find that the diode efficiency approaches perfect diode efficiency as $R$ decreases.

\begin{figure}[t]
\centering
\includegraphics[width=0.4\linewidth]{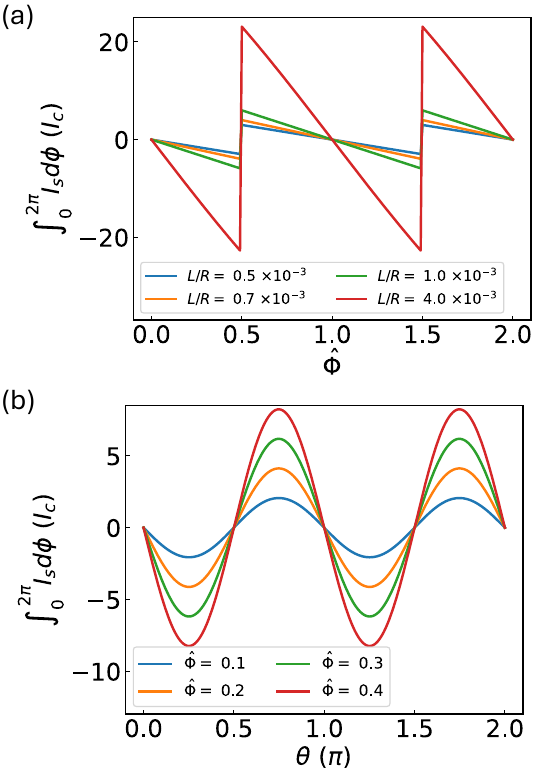}
\caption{\label{fig:fig3}{\bf Non-reciprocal persistent supercurrent:}
Phase-integrated CPR over a $2\pi$ cycle versus (a) $\hat{\Phi}$ for various ratios of $L/R$, and
(b) $\theta$ for $\hat{\Phi} = 0.1,~0.2,~0.3,~0.4$ using the normalization $I_c = \frac{4e\hbar \rho_1 \psi_{\infty}^2 A_{\perp}}{L}$.
Here we used numerically solved Eq.~(\ref{eq:Jnt_2}-\ref{eq:GLnt_2}) as described in the main text with $m_1/m_0 = 100$, $\frac{\hbar^2}{m_0 \zeta_0 L^2} = 2 \times 10^{-3}$, $\frac{\hbar^2 m_0}{m_1^2 \zeta_0 L^2} = \frac{\hbar^2 m_0}{m_1^2 \zeta_0 L^2} = 2 \times 10^{-4}$, and $\frac{\alpha m_0 L^2}{\hbar^2} = 10^{-4}$.}
\end{figure}

In Fig.~\ref{fig:fig2}(b) we present $\eta$ of the ChNt-WL as a function of the chiral angle $\theta$ of the nanotube. Here we see $\eta$ is $\pi$-periodic in $\theta$, reflecting the $C_2$ rotational symmetry of the GL free energy functional, Eq.~(\ref{eq:GLnt_2}). The suppression of $\theta$ at $\pi/2$ and $3\pi/2$ arises from the suppression of $\rho_3 \propto \sin 2\theta$ in the superfluid stiffness tensor. Figure~\ref{fig:fig2}(c) shows $\eta$ versus $m_1$ (which controls the superfluid stiffness anisotropy to leading order) for $\hat{\Phi}= 0.1$. In the limit $m_1 \gg m_0$, the superfluid stiffness becomes nearly isotropic and $\eta \rightarrow 0$. In the opposite limit $m_1 \lesssim m_0$, $\eta$ approaches perfect diode efficiency regardless of nanotube radius. Importantly, this behavior of $\eta$ occurs despite the small values of $\rho_4,~\rho_6,~\rho_7$ and $\rho_8$ used in the calculations. This is significant since in the absence of higher-order momentum terms in the free energy, the diode effect is completely suppressed. Here we find that even with small contributions from higher-order terms, large anisotropy in lowest-order kinetic terms can induce a large SDE. In Fig.~\ref{fig:fig2}(d), we present the CPR of the ChNt-WL for various $L/R$ ratios and with $\hat{\Phi}=0.1$ and $\theta = 0.1\pi$. As $L/R$ increases, we observe the combination of a phase shift and current offset in $I_s$ such that a diode effect emerges but the phase that minimizes the free energy of the ChNt-JJ remains near zero. While phase shifts in the CPR of JJs when both time-reversal and inversion symmetries are broken have been reported before~\cite{Buzdin2008, Mayer2020}, the current offset indicates there is an unexpected persistent current in the weak link. To investigate this further, in Fig.~\ref{fig:fig3} we calculate the phase-integrated CPR $\int_{0}^{2\pi} I_s(\phi)\;d\phi$ which is non-zero when a persistent current is present in the weak link. In Fig.~\ref{fig:fig3}(a), we see the phase-integrated CPR is non-zero when $\eta$ is non-zero (c.f. Fig.~\ref{fig:fig2}(a)) and is linearly proportional to $\hat{\Phi}$. Similarly, we see in Fig.~\ref{fig:fig3}(b) that the persistent current is roughly proportional to $\eta$. This suggests the persistent current is crucial to achieving a near-perfect supercurrent diode efficiency in a ChNt-WL. In the following section, we analytically analyze the origin of the non-reciprocal persistent current in the weak link.

\subsection{Small nanotube diameter limit}
To develop a clear picture for the SDE in ChNt-WLs, we consider the simpler case of a small diameter nanotube where we restrict ourselves to the lowest sub-band for simplicity, and henceforth set $p_y^{(n)} = p_y^{(0)}$. Then the order parameter is
\begin{align}
    \psi(x,y) & = \psi_x(x) e^{ip_y^{(0)} y/\hbar},
\end{align}
for some complex-valued function $\psi_x$.
We can solve Eq.~(\ref{eq:GLnt_2}) by linearizing in the standard way, ignoring higher-order gradients along the axial direction as before, and find
\begin{align}
    \psi_x(x) & \approx \psi_{\infty} L^{-1} e^{iax} \left( L - x + x e^{i(\phi - \phi_0)} \right), \label{eq:psi_short}
\end{align}
where $
\phi_0 = -\frac{\rho_3 p_y^{(0)} L}{\hbar \rho_1}
$.
Then the current-phase relationship of the ChNt-WL is calculated by solving for $J_x$ in Eq.~(\ref{eq:Jnt_2}) which, due to the anisotropic superfluid stiffness, requires both components of $\bf j$:
\begin{align}
    \bf{j} & = 
    \begin{pmatrix}
        2 i \frac{\psi_{\infty}^2}{L} \sin\left(\phi - \phi_0\right) + 2 i \frac{\psi_0^2 \phi_0}{L} \absval{L - x + x e^{i(\phi - \phi_0)}}^2 \\
        \frac{2i \psi_0^2 p_y^{(0)}}{\hbar} \absval{L - x + x e^{i(\phi - \phi_0)}}^2
    \end{pmatrix}.
    \label{eq:jy}
\end{align}
Writing $j_0 = \rho_1 j_x + \rho_3 j_y$ and noting $\partial_x j_0 = 0 = \partial_y j_0$, we can re-arrange the expression for $J_x$ in Eq.~(\ref{eq:Jnt_2}) to be
\begin{align}
    & J_x = \frac{e\hbar}{i\rho_1} \left[2 \rho_1 j_0 -  (\rho_8 \rho_1 - 2\rho_3 \rho_4)p_x^2 j_y \right] \nonumber \\
     + & \frac{e\hbar}{i\rho_1} \left[(\rho_8 \rho_1 + 2\rho_3 \rho_7)p_y^2 + (\rho_6 \rho_1 - 2 \rho_3 \rho_8) p_x p_y \right] j_y.
\end{align}
From Eq.~(\ref{eq:jy}), we see that $p_y j_y = 0$. Then the CPR is given by
\begin{align}
    I_s(\phi) & = I_c \left[\sin \tilde{\phi} + \frac{2\hat{\Phi}}{LR} \gamma^{-1} \left( 1 - \cos \tilde{\phi} \right) \right]. \label{eq:CPR_diode}
\end{align}
Here $I_c = \frac{4e\hbar \rho_1 \psi_{\infty}^2 A_{\perp}}{L}$, $\tilde{\phi} = \phi - \phi_0$,
$\hat{\Phi} = \pi R^2 B_{ext}/\Phi_0$, and
\begin{align}
    \gamma & = \frac{(\frac{m_1}{m_0} + \cos 2\theta)^2 \csc 2\theta}{2 m_1  (\kappa_1 - \lambda - 2\lambda \frac{m_1}{m_0} \cos 2\theta)}.
\end{align}
The CPR in Eq.~(\ref{eq:CPR_diode}) is invariant under a $2\pi$-phase shift $I_s(\phi) = I_s(\phi+2\pi)$, reflecting the fact that changing the phase of the order parameter in either of the leads by $2\pi$ does not change the physical state. The critical current $I_c$ is distinguished from the conventional expression by the chiral angle dependence of $\rho_1 = \rho_1(\theta)$. 

Let's first consider the simple case where $\mathcal{O}(p^4)$ terms in Eq.~(\ref{eq:nt_F}) vanish ($\gamma^{-1} = 0$) so that $I_s = I_c \sin \tilde{\phi}$.
Here we have an anomalous phase $\phi_0$
where the ChNt-WL has the minimum of its free energy at $\phi_0 \ne 0$ and $I_s(-\phi) \ne -I_s(\phi)$ (allowed under broken time-reversal symmetry).
In terms of the anisotropy of the nanotube,
$
\phi_0\propto \sin 2\theta / m_1,
$
and in geometry and magnetic field $\phi_0 \propto B_{ext} A_n$ where $A_n$ is the surface area of the nanotube in the junction. The latter relationship implies $\phi_0$ is linear in the junction length $L$, similar to short Rashba junctions~\cite{Buzdin2008}. 
The presence of an anomalous phase here is due to a purely orbital mechanism~\cite{Banerjee2023} rather than the more common spin-orbit mechanism~\cite{Buzdin2008, Mayer2020}.

We can also gain some intuition about the anomalous phase $\phi_0$ from an analysis of the velocity $v(p_x)$. In this case, we have $\Delta v_x = v_x(p_x) - v_x(-p_x) = 4\rho_3 p_y^{(0)}$ so that
\begin{align}
    \phi_0 & = -\frac{\Delta v_x L}{4\hbar \rho_1}.
\end{align}
Prior work on a Rashba nanowire JJ with a magnetic field perpendicular to the current flow showed that the anomalous phase due to the spin-orbit interaction~\cite{Yokoyama2014} is related to a phase shift
$
\varphi_0 \propto \Delta v_{F,\sigma} L
$
in the Andreev bound state spectrum, where $\Delta v_{F,\sigma}$ is the difference in Fermi velocities of the two spin channels in the junction.
Thus, the anomalous phases in Rashba JJs and ChNt-WLs are directly related to broken chiral symmetry in the condensate velocity.
This suggests an anomalous phase can arise when the condensate velocity is non-reciprocal, and the JDE develops when the non-reciprocity in the condensate velocity cannot be gauged away in the condensate wavefunction~\cite{Hasan2024}.
        
\begin{figure*}[t]
\centering
\includegraphics[width=0.98\linewidth]{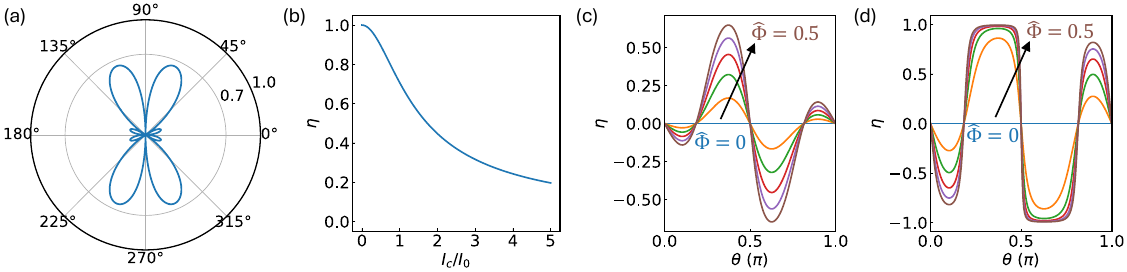}
\caption{\label{fig:diode}{\bf Optimizing diode efficiency:}
(a) $\vert \eta \vert$ versus $\theta$ in polar coordinates for $\hat{\Phi}=0.5$ and $L/R = 10$.
(b) $\eta$ versus $I_0 / I_c$.
$\eta$ versus $\theta$ for $\hat{\Phi}=0,~0.1,...,~0.5$ with (c) $L/R = 10$ and (d) $L/R = 1$.
Here we used $m_1/m_0 = 5$, $\kappa_1 m_1 / R^2  = 50$, and $\lambda m_1 / R^2 = 10$ in Eq.~(\ref{eq:diode_eff}).
}
\end{figure*}

Returning to the full solution in Eq.~(\ref{eq:CPR_diode}), we observe the CPR takes an unconventional bipartite form $I_s(\phi) = \tilde{I}_s(\phi) + I_{0}$ where $I_{0}$ is \textit{independent} of $\phi$. The phase-independent term $I_0$ represents a persistent current in the chiral tube associated with the supercurrent flowing around the tube due to the Little-Parks effect. We have $I_0 \propto B_{ext} \sin(2\theta)$ so that the current is non-zero only when the tube is chiral and an external magnetic field is applied. While a magnetic field-induced supercurrent around the tube (i.e. $J_y \ne 0$) is conventionally expected, it is surprising that this persistent current contributes to the diode effect directly. The picture for the persistent current in this case is depicted in Fig.~\ref{fig:setup}(b), where the current flows along a \textit{helical} path around the chiral tube. 
To preserve the requirement that persistent currents form closed loops in equilibrium, 
the junction will self-tune to the same anomalous phase ($\phi = \phi_0$) so that no net supercurrent flows along the $x$-direction. 

We can compare the persistent current in Eq.~(\ref{eq:CPR_diode}) to the phase-independent supercurrent in the CPR of a junction with Meissner screening inducing non-reciprocity~\cite{Davydova2022}. In Ref.~\onlinecite{Davydova2022}, the Meissner effect generates a persistent current due to spectral flow of Andreev bound states that leads to $\int_0^{2\pi} I_s d\phi = 0$, but in our work we find $\int_0^{2\pi} I_s d\phi \ne 0$. 
The screening current in Ref.~\onlinecite{Davydova2022} is found to induce a diode effect only when higher harmonics enter the CPR i.e. the diode effect vanishes in the CPR to second order in the tunneling amplitude. This suggests the persistent current generated by the Meissner effect in the superconducting electrodes enters higher-order pair tunneling channels across the junction, leading to an interference between channels that causes a diode effect. 
In our case, the persistent current leading to a diode effect is not dependent on pair channel interference or any phase-coherent Josephson tunneling process. The persistent current is associated with $\mathcal{O}(p^4)$ terms in Eq.~(\ref{eq:nt_F}), and we found replacing $\mathcal{O}(p^4)$ terms with terms describing pair co-tunneling does not necessarily result in a diode effect (see the SI). Furthermore, the persistent current is protected by fluxoid quantization so that when the order parameter $\psi$ is non-zero in the ChNt-WL, the non-reciprocal component of the persistent current generally flows across the junction. 
A similar persistent current is also found in asymmetric SQUIDs~\cite{Cuozzo2024_squid}. There, fluxoid quantization dictates the magnetic properties of the device and screening introduces a non-reciprocal persistent current contributing to the supercurrent diode effect, but pair co-tunneling is also found to be necessary for the diode effect~\cite{Cuozzo2024_squid}. Thus, ChNt-WLs exhibit a new type of SDE with persistent currents where pair co-tunneling is unnecessary.

Now let's characterize the diode effect using Eq.~(\ref{eq:CPR_diode}). The diode efficiency takes a particularly simple form:
$
    \eta = I_0/\tilde{I}_c,
$
where $\tilde{I}_c$ is the maximum of $\tilde{I}_s(\phi)$. This can be solved analytically:
\begin{align}
    \eta & = \mathrm{sgn}(\gamma) \frac{2 \hat{\Phi}/(LR)}{\sqrt{ (2 \hat{\Phi}/(LR))^2 + \gamma^2}},
    \label{eq:diode_eff}
\end{align}
Here we assumed $\gamma^{-1} \ne 0$; otherwise, the diode effect vanishes ($\eta = 0$). A representative calculation of $\eta$ as a function of the chiral angle $\theta$ is shown in Fig.~\ref{fig:diode}(a) showing $\eta$ is suppressed at chiral angles $\theta = n\pi/2$ for $n\in \mathbb{Z}$, consistent with our numerical calculations and Ref.~\onlinecite{He2023}. We also observe suppression of $\eta$ at some chiral angles $\theta_0$ (e.g. $\sim 0.18\pi$) where $\gamma^{-1} = 0$ for $\cos 2\theta_0 = \frac{m_0 (\kappa_1 - \lambda)}{2\lambda m_1}$.
 
We observe that $\eta$ is \textit{independent} of temperature.
This is due to the two quantities $(I_{c+}+I_{c-})$ and $(I_{c+} - I_{c-})$ having the same temperature scaling. 
This stands in contrast to the superconducting diode effect predicted using the GL theory for a superconducting chiral nanotube~\cite{He2023} (i.e. in the absence of the junction) where the $\eta$ in that case is sensitive to $T$. In Ref.~\onlinecite{He2023}, the diode effect is calculated using a phenomenological pair breaking Cooper pair momentum along the nanotube and results in a small $\eta$ ($<0.03$). The result in Eq.~(\ref{eq:diode_eff}) ignores a depairing momentum in the superconducting leads. This pair breaking mechanism is expected to play some role in actual experiments, suggesting the diode effect may have some weak $T$-dependence in reality.

\section{Discussion}
Evaluating the upper bounds on $\eta$ we see $\eta \rightarrow 1$ as 
$
I_c/I_0 = \gamma L R / 2\hat{\Phi} \rightarrow 0,
$
see Fig.~\ref{fig:diode}(b).
This limit is, in principle, achieved by optimizing the extrinsic contribution $\hat{\Phi}/LR$ to dominate over the intrinsic contribution $\gamma$ determined by the superfluid stiffness of the chiral nanotube (i.e. $m_1,m_2,\kappa_1,\lambda$). Here $p_y^{(0)}$ cannot be arbitrarily enhanced by decreasing $R$ since a larger kinetic energy of the condensate eventually leads to a suppression of the superconducting state~\cite{He2023}. Then the key parameter for optimizing $\eta$ is a small junction length $L$, as shown in Fig.~\ref{fig:diode}(c-d). Figure~\ref{fig:diode}(c) presents $\eta$ versus $\theta$ for a moderate junction length $L/R = 10$. Here $\eta$ achieves maximum values gradually approaching $0.7$ when $\hat{\Phi}$ is increased to $0.5$. However, when $L/R = 1$, Fig.~\ref{fig:diode}(d) shows that $\eta$ quickly approaches nearly perfect diode efficiency ($\eta = 1$) as $\hat{\Phi}$ increases beyond $0.2$. 
We quantify the dominance of this extrinsic contribution in terms of a geometric quantity $\eta = \sin \Theta$ where $\Theta = \arctan(I_0 / I_c)$, which implies $\eta$ is bounded to values $\vert \eta \vert \le 1$. Thus, in principle it is possible to achieve perfect diode efficiency without non-equilibrium effects~\cite{Souto2024, Su2024, Valentini2024}. Practically speaking, $\eta$ can only approach the ideal limit since $\frac{m_1}{m_0} + \cos 2\theta > 0$ in this chiral nanotube system since we assumed $m_0 < m_1$ (as is typically the case), but there is not a fundamental restriction against $m_0 = m_1$ where we also found numerically in Fig.~\ref{fig:fig2}(c) that $\eta$ approaches unity.
The key ingredients we identify in our work to realize perfect SDE in a weak link are: (i) an anisotropic superfluid stiffness and (ii) fluxoid quantization associated with the applied axial magnetic field which gives rise to a non-reciprocal persistent current across the JJ. The existence of a non-reciprocal persistent current in the JJ is not constrained to the vicinity of the superconducting critical temperature; thus, we expect our analysis to also be relevant at low temperatures.

\section{Conclusion}
In this work, we presented a GL theory for a ChNt-WL. We derived a purely orbital anomalous phase that develops across the junction when a magnetic field is applied parallel to the tube.
The diode effect is suspected to be intimately connected with the anomalous phase~\cite{Amundsen2024}, but here we find it to be independent of the SDE.
We have also shown the origin of the diode effect here is a non-reciprocal persistent current that is protected by fluxoid quantization and can lead to a nearly perfect diode efficiency.
We found this non-reciprocal persistent current can lead to perfect SDE in a weak link—a counter-intuitive result in the context of the basic theory of the dc Josephson effect where persistent currents cannot contribute to the Josephson energy.
In our analysis, we neglected the effect of thermally excited quasiparticles because the key ingredients we identify in our work to realize perfect SDE in a JJ are unchanged: (i) an anisotropic superfluid stiffness and (ii) fluxoid quantization associated with the applied axial magnetic field which gives rise to a non-reciprocal persistent current across the JJ. The presence of thermally excited Bogoliubov quasiparticles will certainly affect the voltage-current characteristic of the weak link, but a central result of our work is that nearly perfect SDE can be achieved if the extrinsic non-reciprocal persistent current dominates over the intrinsic Josephson current in the junction.
The phase-independent persistent current arises specifically in a geometry where fluxoid quantization is a constraint on the gauge-invariant superconducting phase and a combination of mirror and time-reversal symmetries are broken, such as in asymmetric SQUIDs with a finite self-inductance.

We want to emphasize that the key finding in our work is a resolution to the problem of achieving perfect diode efficiency in a Josephson system. In our work, we overcome this challenge by showing a persistent current is found in the chiral nanotube weak link. Despite starting with the same free energy functional as Ref.~{\onlinecite{He2023}}, our calculation is distinct from that of Ref.~{\onlinecite{He2023}} where they calculate extrema of the supercurrent density with respect to variations in a phenomenological depairing momentum. In our approach, we calculate extrema of the CPR, which does not consider a depairing momentum. The superconducting diode efficiency reported in Ref.~{\onlinecite{He2023}} has a maximum value less than 0.03. Using our model for a weak link with comparable parameters, we find a nearly perfect diode efficiency ($>$0.99). This is the first step in developing a theoretical understanding of a perfect dc SDE in a chiral nanotube-based Josephson junction, paving the way for future experiments in this direction.
The mechanism for approaching perfect SDE is also distinguished from prior work in the literature on spin-orbit coupling-based mechanisms. At the level of Ginzburg-Landau theory, spin-orbit coupling-based mechanisms arise from a Lifshitz invariant in the free energy density to account for broken spatial inversion symmetry~{\cite{Kochan2023, Hasan2024}}. In contrast, in our work the free energy density is invariant under spatial inversion. Secondly, while both chiral nanotube and Rashba JJs have a broken mirror symmetry across the junction, the SDE in Rashba systems is usually described in terms of the Zeeman effect while the SDE in the chiral nanotube is associated with an orbital magnetic effect. A consequence of gauge invariance in a nanotube geometry is fluxoid quantization (e.g. Little-Parks effect) which is crucial for nearly perfect diode efficiency in our work. A theoretical proposal to use a persistent current to achieve perfect diode efficiency in a proximity-induced superconducting material with band asymmetry was reported previously~{\cite{Hosur2023}}. We show a non-reciprocal persistent current can develop in a weak link and lead to perfect SDE, opening the door to further study of Josephson devices where unusual persistent currents dictate the non-reciprocal transport properties.

While our analysis is not done for a specific material system, the general arguments are applicable to a wide variety of ChNt-WLs.
An ideal setting to test SDE in chiral nanotubes is with single-walled carbon nanotubes.
Long coherence length can be achieved with common conventional superconductors like Al ($\sim$1 um). Devices with relatively low trapped charge disorder can be made by suspending carbon nanotubes across a trench~{\cite{Cubaynes2020}} or by encapsulating tubes with a hexagonal boron nitride thin film~{\cite{Bauml2021}}, similar to graphene-based devices. In the case of a single-walled carbon nanotube, an axial magnetic field has minimal magnetic flux penetrating the tube so that effects due to vortices are minimal.
It has been demonstrated that a thin flake of NbSe$_2$ can induce superconductivity in carbon nanotubes and enable superconducting effects to be probed at high fields~\cite{Bauml2021}. This could be a useful platform for future studies of supercurrent diode effects in chiral nanotubes.
\\

\paragraph*{Acknowledgements}
J.J.C thanks Catalin D. Spataru, Wei Pan and Enrico Rossi for fruitful discussions.
The work at Sandia is supported by a LDRD project and the U.S. Department of Energy, Office of Science, under the Accelerate Innovations in Emerging Technologies Program.
Sandia National Laboratories is a multi-mission laboratory managed and operated by National Technology \& Engineering Solutions of Sandia, LLC (NTESS), a wholly owned subsidiary of Honeywell International Inc., for the U.S. Department of Energy’s National Nuclear Security Administration (DOE/NNSA) under contract DE-NA0003525. This written work is authored by an employee of NTESS. The employee, not NTESS, owns the right, title and interest in and to the written work and is responsible for its contents. Any subjective views or opinions that might be expressed in the written work do not necessarily represent the views of the U.S. Government. The publisher acknowledges that the U.S. Government retains a non-exclusive, paid-up, irrevocable, world-wide license to publish or reproduce the published form of this written work or allow others to do so, for U.S. Government purposes. The DOE will provide public access to results of federally sponsored research in accordance with the DOE Public Access Plan.

\bibliography{refs}   

\begin{thebibliography}{54}%
\makeatletter
\providecommand \@ifxundefined [1]{%
 \@ifx{#1\undefined}
}%
\providecommand \@ifnum [1]{%
 \ifnum #1\expandafter \@firstoftwo
 \else \expandafter \@secondoftwo
 \fi
}%
\providecommand \@ifx [1]{%
 \ifx #1\expandafter \@firstoftwo
 \else \expandafter \@secondoftwo
 \fi
}%
\providecommand \natexlab [1]{#1}%
\providecommand \enquote  [1]{``#1''}%
\providecommand \bibnamefont  [1]{#1}%
\providecommand \bibfnamefont [1]{#1}%
\providecommand \citenamefont [1]{#1}%
\providecommand \href@noop [0]{\@secondoftwo}%
\providecommand \href [0]{\begingroup \@sanitize@url \@href}%
\providecommand \@href[1]{\@@startlink{#1}\@@href}%
\providecommand \@@href[1]{\endgroup#1\@@endlink}%
\providecommand \@sanitize@url [0]{\catcode `\\12\catcode `\$12\catcode `\&12\catcode `\#12\catcode `\^12\catcode `\_12\catcode `\%12\relax}%
\providecommand \@@startlink[1]{}%
\providecommand \@@endlink[0]{}%
\providecommand \url  [0]{\begingroup\@sanitize@url \@url }%
\providecommand \@url [1]{\endgroup\@href {#1}{\urlprefix }}%
\providecommand \urlprefix  [0]{URL }%
\providecommand \Eprint [0]{\href }%
\providecommand \doibase [0]{https://doi.org/}%
\providecommand \selectlanguage [0]{\@gobble}%
\providecommand \bibinfo  [0]{\@secondoftwo}%
\providecommand \bibfield  [0]{\@secondoftwo}%
\providecommand \translation [1]{[#1]}%
\providecommand \BibitemOpen [0]{}%
\providecommand \bibitemStop [0]{}%
\providecommand \bibitemNoStop [0]{.\EOS\space}%
\providecommand \EOS [0]{\spacefactor3000\relax}%
\providecommand \BibitemShut  [1]{\csname bibitem#1\endcsname}%
\let\auto@bib@innerbib\@empty
\bibitem [{\citenamefont {Zhang}\ \emph {et~al.}(2022)\citenamefont {Zhang}, \citenamefont {Gu}, \citenamefont {Li}, \citenamefont {Hu},\ and\ \citenamefont {Jiang}}]{Zhang2022}%
  \BibitemOpen
  \bibfield  {author} {\bibinfo {author} {\bibfnamefont {Y.}~\bibnamefont {Zhang}}, \bibinfo {author} {\bibfnamefont {Y.}~\bibnamefont {Gu}}, \bibinfo {author} {\bibfnamefont {P.}~\bibnamefont {Li}}, \bibinfo {author} {\bibfnamefont {J.}~\bibnamefont {Hu}},\ and\ \bibinfo {author} {\bibfnamefont {K.}~\bibnamefont {Jiang}},\ }\bibfield  {title} {\bibinfo {title} {General theory of josephson diodes},\ }\href {https://doi.org/10.1103/PhysRevX.12.041013} {\bibfield  {journal} {\bibinfo  {journal} {Phys. Rev. X}\ }\textbf {\bibinfo {volume} {12}},\ \bibinfo {pages} {041013} (\bibinfo {year} {2022})}\BibitemShut {NoStop}%
\bibitem [{\citenamefont {Nadeem}\ \emph {et~al.}(2023)\citenamefont {Nadeem}, \citenamefont {Fuhrer},\ and\ \citenamefont {Wang}}]{Nadeem2023}%
  \BibitemOpen
  \bibfield  {author} {\bibinfo {author} {\bibfnamefont {M.}~\bibnamefont {Nadeem}}, \bibinfo {author} {\bibfnamefont {M.~S.}\ \bibnamefont {Fuhrer}},\ and\ \bibinfo {author} {\bibfnamefont {X.}~\bibnamefont {Wang}},\ }\bibfield  {title} {\bibinfo {title} {The superconducting diode effect},\ }\href {https://doi.org/10.1038/s42254-023-00632-w} {\bibfield  {journal} {\bibinfo  {journal} {Nature Reviews Physics}\ }\textbf {\bibinfo {volume} {5}},\ \bibinfo {pages} {558} (\bibinfo {year} {2023})}\BibitemShut {NoStop}%
\bibitem [{\citenamefont {Hu}\ \emph {et~al.}(2007)\citenamefont {Hu}, \citenamefont {Wu},\ and\ \citenamefont {Dai}}]{Hu2007}%
  \BibitemOpen
  \bibfield  {author} {\bibinfo {author} {\bibfnamefont {J.}~\bibnamefont {Hu}}, \bibinfo {author} {\bibfnamefont {C.}~\bibnamefont {Wu}},\ and\ \bibinfo {author} {\bibfnamefont {X.}~\bibnamefont {Dai}},\ }\bibfield  {title} {\bibinfo {title} {Proposed design of a josephson diode},\ }\href {https://doi.org/10.1103/PhysRevLett.99.067004} {\bibfield  {journal} {\bibinfo  {journal} {Phys. Rev. Lett.}\ }\textbf {\bibinfo {volume} {99}},\ \bibinfo {pages} {067004} (\bibinfo {year} {2007})}\BibitemShut {NoStop}%
\bibitem [{\citenamefont {Halterman}\ \emph {et~al.}(2022)\citenamefont {Halterman}, \citenamefont {Alidoust}, \citenamefont {Smith},\ and\ \citenamefont {Starr}}]{Halterman2022}%
  \BibitemOpen
  \bibfield  {author} {\bibinfo {author} {\bibfnamefont {K.}~\bibnamefont {Halterman}}, \bibinfo {author} {\bibfnamefont {M.}~\bibnamefont {Alidoust}}, \bibinfo {author} {\bibfnamefont {R.}~\bibnamefont {Smith}},\ and\ \bibinfo {author} {\bibfnamefont {S.}~\bibnamefont {Starr}},\ }\bibfield  {title} {\bibinfo {title} {Supercurrent diode effect, spin torques, and robust zero-energy peak in planar half-metallic trilayers},\ }\href {https://doi.org/10.1103/PhysRevB.105.104508} {\bibfield  {journal} {\bibinfo  {journal} {Phys. Rev. B}\ }\textbf {\bibinfo {volume} {105}},\ \bibinfo {pages} {104508} (\bibinfo {year} {2022})}\BibitemShut {NoStop}%
\bibitem [{\citenamefont {Wu}\ \emph {et~al.}(2022{\natexlab{a}})\citenamefont {Wu}, \citenamefont {Wang}, \citenamefont {Xu}, \citenamefont {Sivakumar}, \citenamefont {Pasco}, \citenamefont {Filippozzi}, \citenamefont {Parkin}, \citenamefont {Zeng}, \citenamefont {McQueen},\ and\ \citenamefont {Ali}}]{Wu2022}%
  \BibitemOpen
  \bibfield  {author} {\bibinfo {author} {\bibfnamefont {H.}~\bibnamefont {Wu}}, \bibinfo {author} {\bibfnamefont {Y.}~\bibnamefont {Wang}}, \bibinfo {author} {\bibfnamefont {Y.}~\bibnamefont {Xu}}, \bibinfo {author} {\bibfnamefont {P.~K.}\ \bibnamefont {Sivakumar}}, \bibinfo {author} {\bibfnamefont {C.}~\bibnamefont {Pasco}}, \bibinfo {author} {\bibfnamefont {U.}~\bibnamefont {Filippozzi}}, \bibinfo {author} {\bibfnamefont {S.~S.~P.}\ \bibnamefont {Parkin}}, \bibinfo {author} {\bibfnamefont {Y.-J.}\ \bibnamefont {Zeng}}, \bibinfo {author} {\bibfnamefont {T.}~\bibnamefont {McQueen}},\ and\ \bibinfo {author} {\bibfnamefont {M.~N.}\ \bibnamefont {Ali}},\ }\bibfield  {title} {\bibinfo {title} {The field-free josephson diode in a van der waals heterostructure},\ }\href {https://doi.org/10.1038/s41586-022-04504-8} {\bibfield  {journal} {\bibinfo  {journal} {Nature}\ }\textbf {\bibinfo {volume} {604}},\ \bibinfo {pages} {653} (\bibinfo {year} {2022}{\natexlab{a}})}\BibitemShut {NoStop}%
\bibitem [{\citenamefont {Narita}\ \emph {et~al.}(2022)\citenamefont {Narita}, \citenamefont {Ishizuka}, \citenamefont {Kawarazaki}, \citenamefont {Kan}, \citenamefont {Shiota}, \citenamefont {Moriyama}, \citenamefont {Shimakawa}, \citenamefont {Ognev}, \citenamefont {Samardak}, \citenamefont {Yanase},\ and\ \citenamefont {Ono}}]{narita_field-free_2022}%
  \BibitemOpen
  \bibfield  {author} {\bibinfo {author} {\bibfnamefont {H.}~\bibnamefont {Narita}}, \bibinfo {author} {\bibfnamefont {J.}~\bibnamefont {Ishizuka}}, \bibinfo {author} {\bibfnamefont {R.}~\bibnamefont {Kawarazaki}}, \bibinfo {author} {\bibfnamefont {D.}~\bibnamefont {Kan}}, \bibinfo {author} {\bibfnamefont {Y.}~\bibnamefont {Shiota}}, \bibinfo {author} {\bibfnamefont {T.}~\bibnamefont {Moriyama}}, \bibinfo {author} {\bibfnamefont {Y.}~\bibnamefont {Shimakawa}}, \bibinfo {author} {\bibfnamefont {A.~V.}\ \bibnamefont {Ognev}}, \bibinfo {author} {\bibfnamefont {A.~S.}\ \bibnamefont {Samardak}}, \bibinfo {author} {\bibfnamefont {Y.}~\bibnamefont {Yanase}},\ and\ \bibinfo {author} {\bibfnamefont {T.}~\bibnamefont {Ono}},\ }\bibfield  {title} {\bibinfo {title} {Field-free superconducting diode effect in noncentrosymmetric superconductor/ferromagnet multilayers},\ }\href {https://doi.org/10.1038/s41565-022-01159-4} {\bibfield  {journal} {\bibinfo  {journal} {Nature Nanotechnology}\ }\textbf {\bibinfo {volume} {17}},\
  \bibinfo {pages} {823} (\bibinfo {year} {2022})},\ \bibinfo {note} {publisher: Nature Publishing Group}\BibitemShut {NoStop}%
\bibitem [{\citenamefont {Hou}\ \emph {et~al.}(2023)\citenamefont {Hou}, \citenamefont {Nichele}, \citenamefont {Chi}, \citenamefont {Lodesani}, \citenamefont {Wu}, \citenamefont {Ritter}, \citenamefont {Haxell}, \citenamefont {Davydova}, \citenamefont {Ili\ifmmode~\acute{c}\else \'{c}\fi{}}, \citenamefont {Glezakou-Elbert}, \citenamefont {Varambally}, \citenamefont {Bergeret}, \citenamefont {Kamra}, \citenamefont {Fu}, \citenamefont {Lee},\ and\ \citenamefont {Moodera}}]{Hou2023}%
  \BibitemOpen
  \bibfield  {author} {\bibinfo {author} {\bibfnamefont {Y.}~\bibnamefont {Hou}}, \bibinfo {author} {\bibfnamefont {F.}~\bibnamefont {Nichele}}, \bibinfo {author} {\bibfnamefont {H.}~\bibnamefont {Chi}}, \bibinfo {author} {\bibfnamefont {A.}~\bibnamefont {Lodesani}}, \bibinfo {author} {\bibfnamefont {Y.}~\bibnamefont {Wu}}, \bibinfo {author} {\bibfnamefont {M.~F.}\ \bibnamefont {Ritter}}, \bibinfo {author} {\bibfnamefont {D.~Z.}\ \bibnamefont {Haxell}}, \bibinfo {author} {\bibfnamefont {M.}~\bibnamefont {Davydova}}, \bibinfo {author} {\bibfnamefont {S.}~\bibnamefont {Ili\ifmmode~\acute{c}\else \'{c}\fi{}}}, \bibinfo {author} {\bibfnamefont {O.}~\bibnamefont {Glezakou-Elbert}}, \bibinfo {author} {\bibfnamefont {A.}~\bibnamefont {Varambally}}, \bibinfo {author} {\bibfnamefont {F.~S.}\ \bibnamefont {Bergeret}}, \bibinfo {author} {\bibfnamefont {A.}~\bibnamefont {Kamra}}, \bibinfo {author} {\bibfnamefont {L.}~\bibnamefont {Fu}}, \bibinfo {author} {\bibfnamefont {P.~A.}\ \bibnamefont {Lee}},\ and\ \bibinfo
  {author} {\bibfnamefont {J.~S.}\ \bibnamefont {Moodera}},\ }\bibfield  {title} {\bibinfo {title} {Ubiquitous superconducting diode effect in superconductor thin films},\ }\href {https://doi.org/10.1103/PhysRevLett.131.027001} {\bibfield  {journal} {\bibinfo  {journal} {Phys. Rev. Lett.}\ }\textbf {\bibinfo {volume} {131}},\ \bibinfo {pages} {027001} (\bibinfo {year} {2023})}\BibitemShut {NoStop}%
\bibitem [{\citenamefont {He}\ \emph {et~al.}(2022)\citenamefont {He}, \citenamefont {Tanaka},\ and\ \citenamefont {Nagaosa}}]{He2022}%
  \BibitemOpen
  \bibfield  {author} {\bibinfo {author} {\bibfnamefont {J.~J.}\ \bibnamefont {He}}, \bibinfo {author} {\bibfnamefont {Y.}~\bibnamefont {Tanaka}},\ and\ \bibinfo {author} {\bibfnamefont {N.}~\bibnamefont {Nagaosa}},\ }\bibfield  {title} {\bibinfo {title} {A phenomenological theory of superconductor diodes},\ }\href {https://doi.org/10.1088/1367-2630/ac6766} {\bibfield  {journal} {\bibinfo  {journal} {New Journal of Physics}\ }\textbf {\bibinfo {volume} {24}},\ \bibinfo {pages} {053014} (\bibinfo {year} {2022})}\BibitemShut {NoStop}%
\bibitem [{\citenamefont {He}\ \emph {et~al.}(2023)\citenamefont {He}, \citenamefont {Tanaka},\ and\ \citenamefont {Nagaosa}}]{He2023}%
  \BibitemOpen
  \bibfield  {author} {\bibinfo {author} {\bibfnamefont {J.~J.}\ \bibnamefont {He}}, \bibinfo {author} {\bibfnamefont {Y.}~\bibnamefont {Tanaka}},\ and\ \bibinfo {author} {\bibfnamefont {N.}~\bibnamefont {Nagaosa}},\ }\bibfield  {title} {\bibinfo {title} {The supercurrent diode effect and nonreciprocal paraconductivity due to the chiral structure of nanotubes},\ }\href {https://doi.org/10.1038/s41467-023-39083-3} {\bibfield  {journal} {\bibinfo  {journal} {Nature Communications}\ }\textbf {\bibinfo {volume} {14}},\ \bibinfo {pages} {3330} (\bibinfo {year} {2023})}\BibitemShut {NoStop}%
\bibitem [{\citenamefont {Shi}\ \emph {et~al.}(2015)\citenamefont {Shi}, \citenamefont {Yu}, \citenamefont {Jiang}, \citenamefont {Andrei~Bernevig}, \citenamefont {Pan}, \citenamefont {Hawkins},\ and\ \citenamefont {Klem}}]{Shi2015}%
  \BibitemOpen
  \bibfield  {author} {\bibinfo {author} {\bibfnamefont {X.}~\bibnamefont {Shi}}, \bibinfo {author} {\bibfnamefont {W.}~\bibnamefont {Yu}}, \bibinfo {author} {\bibfnamefont {Z.}~\bibnamefont {Jiang}}, \bibinfo {author} {\bibfnamefont {B.}~\bibnamefont {Andrei~Bernevig}}, \bibinfo {author} {\bibfnamefont {W.}~\bibnamefont {Pan}}, \bibinfo {author} {\bibfnamefont {S.~D.}\ \bibnamefont {Hawkins}},\ and\ \bibinfo {author} {\bibfnamefont {J.~F.}\ \bibnamefont {Klem}},\ }\bibfield  {title} {\bibinfo {title} {Giant supercurrent states in a superconductor-inas/gasb-superconductor junction},\ }\href {https://doi.org/10.1063/1.4932644} {\bibfield  {journal} {\bibinfo  {journal} {Journal of Applied Physics}\ }\textbf {\bibinfo {volume} {118}},\ \bibinfo {pages} {133905} (\bibinfo {year} {2015})},\ \Eprint {https://arxiv.org/abs/https://doi.org/10.1063/1.4932644} {https://doi.org/10.1063/1.4932644} \BibitemShut {NoStop}%
\bibitem [{\citenamefont {Bocquillon}\ and\ \citenamefont {\emph{et~al.}}(2017)}]{Bocquillon2017}%
  \BibitemOpen
  \bibfield  {author} {\bibinfo {author} {\bibfnamefont {E.}~\bibnamefont {Bocquillon}}\ and\ \bibinfo {author} {\bibnamefont {\emph{et~al.}}},\ }\bibfield  {title} {\bibinfo {title} {Gapless {Andreev} bound states in the quantum spin {Hall} insulator {HgTe}},\ }\href@noop {} {\bibfield  {journal} {\bibinfo  {journal} {Nature Nanotech}\ }\textbf {\bibinfo {volume} {12}},\ \bibinfo {pages} {137} (\bibinfo {year} {2017})}\BibitemShut {NoStop}%
\bibitem [{\citenamefont {Pal}\ and\ \citenamefont {Benjamin}(2019)}]{Pal2019}%
  \BibitemOpen
  \bibfield  {author} {\bibinfo {author} {\bibfnamefont {S.}~\bibnamefont {Pal}}\ and\ \bibinfo {author} {\bibfnamefont {C.}~\bibnamefont {Benjamin}},\ }\bibfield  {title} {\bibinfo {title} {Quantized josephson phase battery},\ }\href {https://doi.org/10.1209/0295-5075/126/57002} {\bibfield  {journal} {\bibinfo  {journal} {Europhysics Letters}\ }\textbf {\bibinfo {volume} {126}},\ \bibinfo {pages} {57002} (\bibinfo {year} {2019})}\BibitemShut {NoStop}%
\bibitem [{\citenamefont {Misaki}\ and\ \citenamefont {Nagaosa}(2021)}]{Misaki2021}%
  \BibitemOpen
  \bibfield  {author} {\bibinfo {author} {\bibfnamefont {K.}~\bibnamefont {Misaki}}\ and\ \bibinfo {author} {\bibfnamefont {N.}~\bibnamefont {Nagaosa}},\ }\bibfield  {title} {\bibinfo {title} {Theory of the nonreciprocal josephson effect},\ }\href {https://doi.org/10.1103/PhysRevB.103.245302} {\bibfield  {journal} {\bibinfo  {journal} {Phys. Rev. B}\ }\textbf {\bibinfo {volume} {103}},\ \bibinfo {pages} {245302} (\bibinfo {year} {2021})}\BibitemShut {NoStop}%
\bibitem [{\citenamefont {Baumgartner}\ \emph {et~al.}(2022{\natexlab{a}})\citenamefont {Baumgartner}, \citenamefont {Fuchs}, \citenamefont {Costa}, \citenamefont {Picó-Cortés}, \citenamefont {Reinhardt}, \citenamefont {Gronin}, \citenamefont {Gardner}, \citenamefont {Lindemann}, \citenamefont {Manfra}, \citenamefont {Junior}, \citenamefont {Kochan}, \citenamefont {Fabian}, \citenamefont {Paradiso},\ and\ \citenamefont {Strunk}}]{Baumgartner2022_JPhys}%
  \BibitemOpen
  \bibfield  {author} {\bibinfo {author} {\bibfnamefont {C.}~\bibnamefont {Baumgartner}}, \bibinfo {author} {\bibfnamefont {L.}~\bibnamefont {Fuchs}}, \bibinfo {author} {\bibfnamefont {A.}~\bibnamefont {Costa}}, \bibinfo {author} {\bibfnamefont {J.}~\bibnamefont {Picó-Cortés}}, \bibinfo {author} {\bibfnamefont {S.}~\bibnamefont {Reinhardt}}, \bibinfo {author} {\bibfnamefont {S.}~\bibnamefont {Gronin}}, \bibinfo {author} {\bibfnamefont {G.~C.}\ \bibnamefont {Gardner}}, \bibinfo {author} {\bibfnamefont {T.}~\bibnamefont {Lindemann}}, \bibinfo {author} {\bibfnamefont {M.~J.}\ \bibnamefont {Manfra}}, \bibinfo {author} {\bibfnamefont {P.~E.~F.}\ \bibnamefont {Junior}}, \bibinfo {author} {\bibfnamefont {D.}~\bibnamefont {Kochan}}, \bibinfo {author} {\bibfnamefont {J.}~\bibnamefont {Fabian}}, \bibinfo {author} {\bibfnamefont {N.}~\bibnamefont {Paradiso}},\ and\ \bibinfo {author} {\bibfnamefont {C.}~\bibnamefont {Strunk}},\ }\bibfield  {title} {\bibinfo {title} {Effect of rashba and dresselhaus spin–orbit
  coupling on supercurrent rectification and magnetochiral anisotropy of ballistic josephson junctions},\ }\href {https://doi.org/10.1088/1361-648X/ac4d5e} {\bibfield  {journal} {\bibinfo  {journal} {Journal of Physics: Condensed Matter}\ }\textbf {\bibinfo {volume} {34}},\ \bibinfo {pages} {154005} (\bibinfo {year} {2022}{\natexlab{a}})}\BibitemShut {NoStop}%
\bibitem [{\citenamefont {Baumgartner}\ \emph {et~al.}(2022{\natexlab{b}})\citenamefont {Baumgartner}, \citenamefont {Fuchs}, \citenamefont {Costa}, \citenamefont {Reinhardt}, \citenamefont {Gronin}, \citenamefont {Gardner}, \citenamefont {Lindemann}, \citenamefont {Manfra}, \citenamefont {Faria~Junior}, \citenamefont {Kochan}, \citenamefont {Fabian}, \citenamefont {Paradiso},\ and\ \citenamefont {Strunk}}]{Baumgartner2022_NatNano}%
  \BibitemOpen
  \bibfield  {author} {\bibinfo {author} {\bibfnamefont {C.}~\bibnamefont {Baumgartner}}, \bibinfo {author} {\bibfnamefont {L.}~\bibnamefont {Fuchs}}, \bibinfo {author} {\bibfnamefont {A.}~\bibnamefont {Costa}}, \bibinfo {author} {\bibfnamefont {S.}~\bibnamefont {Reinhardt}}, \bibinfo {author} {\bibfnamefont {S.}~\bibnamefont {Gronin}}, \bibinfo {author} {\bibfnamefont {G.~C.}\ \bibnamefont {Gardner}}, \bibinfo {author} {\bibfnamefont {T.}~\bibnamefont {Lindemann}}, \bibinfo {author} {\bibfnamefont {M.~J.}\ \bibnamefont {Manfra}}, \bibinfo {author} {\bibfnamefont {P.~E.}\ \bibnamefont {Faria~Junior}}, \bibinfo {author} {\bibfnamefont {D.}~\bibnamefont {Kochan}}, \bibinfo {author} {\bibfnamefont {J.}~\bibnamefont {Fabian}}, \bibinfo {author} {\bibfnamefont {N.}~\bibnamefont {Paradiso}},\ and\ \bibinfo {author} {\bibfnamefont {C.}~\bibnamefont {Strunk}},\ }\bibfield  {title} {\bibinfo {title} {Supercurrent rectification and magnetochiral effects in symmetric josephson junctions},\ }\href
  {https://doi.org/10.1038/s41565-021-01009-9} {\bibfield  {journal} {\bibinfo  {journal} {Nature Nanotechnology}\ }\textbf {\bibinfo {volume} {17}},\ \bibinfo {pages} {39} (\bibinfo {year} {2022}{\natexlab{b}})}\BibitemShut {NoStop}%
\bibitem [{\citenamefont {Jeon}\ \emph {et~al.}(2022)\citenamefont {Jeon}, \citenamefont {Kim}, \citenamefont {Yoon}, \citenamefont {Jeon}, \citenamefont {Han}, \citenamefont {Cottet}, \citenamefont {Kontos},\ and\ \citenamefont {Parkin}}]{jeon_zero-field_2022}%
  \BibitemOpen
  \bibfield  {author} {\bibinfo {author} {\bibfnamefont {K.-R.}\ \bibnamefont {Jeon}}, \bibinfo {author} {\bibfnamefont {J.-K.}\ \bibnamefont {Kim}}, \bibinfo {author} {\bibfnamefont {J.}~\bibnamefont {Yoon}}, \bibinfo {author} {\bibfnamefont {J.-C.}\ \bibnamefont {Jeon}}, \bibinfo {author} {\bibfnamefont {H.}~\bibnamefont {Han}}, \bibinfo {author} {\bibfnamefont {A.}~\bibnamefont {Cottet}}, \bibinfo {author} {\bibfnamefont {T.}~\bibnamefont {Kontos}},\ and\ \bibinfo {author} {\bibfnamefont {S.~S.~P.}\ \bibnamefont {Parkin}},\ }\bibfield  {title} {\bibinfo {title} {Zero-field polarity-reversible {Josephson} supercurrent diodes enabled by a proximity-magnetized {Pt} barrier},\ }\href {https://doi.org/10.1038/s41563-022-01300-7} {\bibfield  {journal} {\bibinfo  {journal} {Nature Materials}\ }\textbf {\bibinfo {volume} {21}},\ \bibinfo {pages} {1008} (\bibinfo {year} {2022})},\ \bibinfo {note} {publisher: Nature Publishing Group}\BibitemShut {NoStop}%
\bibitem [{\citenamefont {Pal}\ \emph {et~al.}(2022)\citenamefont {Pal}, \citenamefont {Chakraborty}, \citenamefont {Sivakumar}, \citenamefont {Davydova}, \citenamefont {Gopi}, \citenamefont {Pandeya}, \citenamefont {Krieger}, \citenamefont {Zhang}, \citenamefont {Date}, \citenamefont {Ju}, \citenamefont {Yuan}, \citenamefont {Schröter}, \citenamefont {Fu},\ and\ \citenamefont {Parkin}}]{Pal2022}%
  \BibitemOpen
  \bibfield  {author} {\bibinfo {author} {\bibfnamefont {B.}~\bibnamefont {Pal}}, \bibinfo {author} {\bibfnamefont {A.}~\bibnamefont {Chakraborty}}, \bibinfo {author} {\bibfnamefont {P.~K.}\ \bibnamefont {Sivakumar}}, \bibinfo {author} {\bibfnamefont {M.}~\bibnamefont {Davydova}}, \bibinfo {author} {\bibfnamefont {A.~K.}\ \bibnamefont {Gopi}}, \bibinfo {author} {\bibfnamefont {A.~K.}\ \bibnamefont {Pandeya}}, \bibinfo {author} {\bibfnamefont {J.~A.}\ \bibnamefont {Krieger}}, \bibinfo {author} {\bibfnamefont {Y.}~\bibnamefont {Zhang}}, \bibinfo {author} {\bibfnamefont {M.}~\bibnamefont {Date}}, \bibinfo {author} {\bibfnamefont {S.}~\bibnamefont {Ju}}, \bibinfo {author} {\bibfnamefont {N.}~\bibnamefont {Yuan}}, \bibinfo {author} {\bibfnamefont {N.~B.~M.}\ \bibnamefont {Schröter}}, \bibinfo {author} {\bibfnamefont {L.}~\bibnamefont {Fu}},\ and\ \bibinfo {author} {\bibfnamefont {S.~S.~P.}\ \bibnamefont {Parkin}},\ }\bibfield  {title} {\bibinfo {title} {Josephson diode effect from cooper pair momentum in a
  topological semimetal},\ }\href {https://doi.org/10.1038/s41567-022-01699-5} {\bibfield  {journal} {\bibinfo  {journal} {Nature Physics}\ }\textbf {\bibinfo {volume} {18}},\ \bibinfo {pages} {1228} (\bibinfo {year} {2022})}\BibitemShut {NoStop}%
\bibitem [{\citenamefont {Kokkeler}\ \emph {et~al.}(2022)\citenamefont {Kokkeler}, \citenamefont {Golubov},\ and\ \citenamefont {Bergeret}}]{Kokkeler2022}%
  \BibitemOpen
  \bibfield  {author} {\bibinfo {author} {\bibfnamefont {T.~H.}\ \bibnamefont {Kokkeler}}, \bibinfo {author} {\bibfnamefont {A.~A.}\ \bibnamefont {Golubov}},\ and\ \bibinfo {author} {\bibfnamefont {F.~S.}\ \bibnamefont {Bergeret}},\ }\bibfield  {title} {\bibinfo {title} {Field-free anomalous junction and superconducting diode effect in spin-split superconductor/topological insulator junctions},\ }\href {https://doi.org/10.1103/PhysRevB.106.214504} {\bibfield  {journal} {\bibinfo  {journal} {Phys. Rev. B}\ }\textbf {\bibinfo {volume} {106}},\ \bibinfo {pages} {214504} (\bibinfo {year} {2022})}\BibitemShut {NoStop}%
\bibitem [{\citenamefont {Davydova}\ \emph {et~al.}(2022)\citenamefont {Davydova}, \citenamefont {Prembabu},\ and\ \citenamefont {Fu}}]{Davydova2022}%
  \BibitemOpen
  \bibfield  {author} {\bibinfo {author} {\bibfnamefont {M.}~\bibnamefont {Davydova}}, \bibinfo {author} {\bibfnamefont {S.}~\bibnamefont {Prembabu}},\ and\ \bibinfo {author} {\bibfnamefont {L.}~\bibnamefont {Fu}},\ }\bibfield  {title} {\bibinfo {title} {Universal josephson diode effect},\ }\href {https://doi.org/10.1126/sciadv.abo0309} {\bibfield  {journal} {\bibinfo  {journal} {Science Advances}\ }\textbf {\bibinfo {volume} {8}},\ \bibinfo {pages} {eabo0309} (\bibinfo {year} {2022})},\ \Eprint {https://arxiv.org/abs/https://www.science.org/doi/pdf/10.1126/sciadv.abo0309} {https://www.science.org/doi/pdf/10.1126/sciadv.abo0309} \BibitemShut {NoStop}%
\bibitem [{\citenamefont {Ili\ifmmode~\acute{c}\else \'{c}\fi{}}\ \emph {et~al.}(2022)\citenamefont {Ili\ifmmode~\acute{c}\else \'{c}\fi{}}, \citenamefont {Virtanen}, \citenamefont {Heikkil\"a},\ and\ \citenamefont {Bergeret}}]{Illic2022_PRApplied}%
  \BibitemOpen
  \bibfield  {author} {\bibinfo {author} {\bibfnamefont {S.}~\bibnamefont {Ili\ifmmode~\acute{c}\else \'{c}\fi{}}}, \bibinfo {author} {\bibfnamefont {P.}~\bibnamefont {Virtanen}}, \bibinfo {author} {\bibfnamefont {T.~T.}\ \bibnamefont {Heikkil\"a}},\ and\ \bibinfo {author} {\bibfnamefont {F.~S.}\ \bibnamefont {Bergeret}},\ }\bibfield  {title} {\bibinfo {title} {Current rectification in junctions with spin-split superconductors},\ }\href {https://doi.org/10.1103/PhysRevApplied.17.034049} {\bibfield  {journal} {\bibinfo  {journal} {Phys. Rev. Appl.}\ }\textbf {\bibinfo {volume} {17}},\ \bibinfo {pages} {034049} (\bibinfo {year} {2022})}\BibitemShut {NoStop}%
\bibitem [{\citenamefont {Tanaka}\ \emph {et~al.}(2022)\citenamefont {Tanaka}, \citenamefont {Lu},\ and\ \citenamefont {Nagaosa}}]{Tanaka2022}%
  \BibitemOpen
  \bibfield  {author} {\bibinfo {author} {\bibfnamefont {Y.}~\bibnamefont {Tanaka}}, \bibinfo {author} {\bibfnamefont {B.}~\bibnamefont {Lu}},\ and\ \bibinfo {author} {\bibfnamefont {N.}~\bibnamefont {Nagaosa}},\ }\bibfield  {title} {\bibinfo {title} {Theory of giant diode effect in $d$-wave superconductor junctions on the surface of a topological insulator},\ }\href {https://doi.org/10.1103/PhysRevB.106.214524} {\bibfield  {journal} {\bibinfo  {journal} {Phys. Rev. B}\ }\textbf {\bibinfo {volume} {106}},\ \bibinfo {pages} {214524} (\bibinfo {year} {2022})}\BibitemShut {NoStop}%
\bibitem [{\citenamefont {Trahms}\ \emph {et~al.}(2023)\citenamefont {Trahms}, \citenamefont {Melischek}, \citenamefont {Steiner}, \citenamefont {Mahendru}, \citenamefont {Tamir}, \citenamefont {Bogdanoff}, \citenamefont {Peters}, \citenamefont {Reecht}, \citenamefont {Winkelmann}, \citenamefont {von Oppen},\ and\ \citenamefont {Franke}}]{Trahms2023}%
  \BibitemOpen
  \bibfield  {author} {\bibinfo {author} {\bibfnamefont {M.}~\bibnamefont {Trahms}}, \bibinfo {author} {\bibfnamefont {L.}~\bibnamefont {Melischek}}, \bibinfo {author} {\bibfnamefont {J.~F.}\ \bibnamefont {Steiner}}, \bibinfo {author} {\bibfnamefont {B.}~\bibnamefont {Mahendru}}, \bibinfo {author} {\bibfnamefont {I.}~\bibnamefont {Tamir}}, \bibinfo {author} {\bibfnamefont {N.}~\bibnamefont {Bogdanoff}}, \bibinfo {author} {\bibfnamefont {O.}~\bibnamefont {Peters}}, \bibinfo {author} {\bibfnamefont {G.}~\bibnamefont {Reecht}}, \bibinfo {author} {\bibfnamefont {C.~B.}\ \bibnamefont {Winkelmann}}, \bibinfo {author} {\bibfnamefont {F.}~\bibnamefont {von Oppen}},\ and\ \bibinfo {author} {\bibfnamefont {K.~J.}\ \bibnamefont {Franke}},\ }\bibfield  {title} {\bibinfo {title} {Diode effect in josephson junctions with a single magnetic atom},\ }\href {https://doi.org/10.1038/s41586-023-05743-z} {\bibfield  {journal} {\bibinfo  {journal} {Nature}\ }\textbf {\bibinfo {volume} {615}},\ \bibinfo {pages} {628} (\bibinfo
  {year} {2023})}\BibitemShut {NoStop}%
\bibitem [{\citenamefont {Cayao}\ \emph {et~al.}(2024)\citenamefont {Cayao}, \citenamefont {Nagaosa},\ and\ \citenamefont {Tanaka}}]{Cayao2024}%
  \BibitemOpen
  \bibfield  {author} {\bibinfo {author} {\bibfnamefont {J.}~\bibnamefont {Cayao}}, \bibinfo {author} {\bibfnamefont {N.}~\bibnamefont {Nagaosa}},\ and\ \bibinfo {author} {\bibfnamefont {Y.}~\bibnamefont {Tanaka}},\ }\bibfield  {title} {\bibinfo {title} {Enhancing the josephson diode effect with majorana bound states},\ }\href {https://doi.org/10.1103/PhysRevB.109.L081405} {\bibfield  {journal} {\bibinfo  {journal} {Phys. Rev. B}\ }\textbf {\bibinfo {volume} {109}},\ \bibinfo {pages} {L081405} (\bibinfo {year} {2024})}\BibitemShut {NoStop}%
\bibitem [{\citenamefont {Yerin}\ \emph {et~al.}(2024)\citenamefont {Yerin}, \citenamefont {Drechsler}, \citenamefont {Varlamov}, \citenamefont {Cuoco},\ and\ \citenamefont {Giazotto}}]{Yerin2024}%
  \BibitemOpen
  \bibfield  {author} {\bibinfo {author} {\bibfnamefont {Y.}~\bibnamefont {Yerin}}, \bibinfo {author} {\bibfnamefont {S.-L.}\ \bibnamefont {Drechsler}}, \bibinfo {author} {\bibfnamefont {A.~A.}\ \bibnamefont {Varlamov}}, \bibinfo {author} {\bibfnamefont {M.}~\bibnamefont {Cuoco}},\ and\ \bibinfo {author} {\bibfnamefont {F.}~\bibnamefont {Giazotto}},\ }\bibfield  {title} {\bibinfo {title} {Supercurrent rectification with time-reversal symmetry broken multiband superconductors},\ }\href {https://doi.org/10.1103/PhysRevB.110.054501} {\bibfield  {journal} {\bibinfo  {journal} {Phys. Rev. B}\ }\textbf {\bibinfo {volume} {110}},\ \bibinfo {pages} {054501} (\bibinfo {year} {2024})}\BibitemShut {NoStop}%
\bibitem [{\citenamefont {Lu}\ \emph {et~al.}(2023)\citenamefont {Lu}, \citenamefont {Ikegaya}, \citenamefont {Burset}, \citenamefont {Tanaka},\ and\ \citenamefont {Nagaosa}}]{Lu2023}%
  \BibitemOpen
  \bibfield  {author} {\bibinfo {author} {\bibfnamefont {B.}~\bibnamefont {Lu}}, \bibinfo {author} {\bibfnamefont {S.}~\bibnamefont {Ikegaya}}, \bibinfo {author} {\bibfnamefont {P.}~\bibnamefont {Burset}}, \bibinfo {author} {\bibfnamefont {Y.}~\bibnamefont {Tanaka}},\ and\ \bibinfo {author} {\bibfnamefont {N.}~\bibnamefont {Nagaosa}},\ }\bibfield  {title} {\bibinfo {title} {Tunable josephson diode effect on the surface of topological insulators},\ }\href {https://doi.org/10.1103/PhysRevLett.131.096001} {\bibfield  {journal} {\bibinfo  {journal} {Phys. Rev. Lett.}\ }\textbf {\bibinfo {volume} {131}},\ \bibinfo {pages} {096001} (\bibinfo {year} {2023})}\BibitemShut {NoStop}%
\bibitem [{\citenamefont {Debnath}\ and\ \citenamefont {Dutta}(2024)}]{Debnath2024}%
  \BibitemOpen
  \bibfield  {author} {\bibinfo {author} {\bibfnamefont {D.}~\bibnamefont {Debnath}}\ and\ \bibinfo {author} {\bibfnamefont {P.}~\bibnamefont {Dutta}},\ }\bibfield  {title} {\bibinfo {title} {Gate-tunable josephson diode effect in rashba spin-orbit coupled quantum dot junctions},\ }\href {https://doi.org/10.1103/PhysRevB.109.174511} {\bibfield  {journal} {\bibinfo  {journal} {Phys. Rev. B}\ }\textbf {\bibinfo {volume} {109}},\ \bibinfo {pages} {174511} (\bibinfo {year} {2024})}\BibitemShut {NoStop}%
\bibitem [{\citenamefont {Debnath}\ and\ \citenamefont {Dutta}(2025)}]{Debnath2025}%
  \BibitemOpen
  \bibfield  {author} {\bibinfo {author} {\bibfnamefont {D.}~\bibnamefont {Debnath}}\ and\ \bibinfo {author} {\bibfnamefont {P.}~\bibnamefont {Dutta}},\ }\bibfield  {title} {\bibinfo {title} {Field-free josephson diode effect in interacting chiral quantum dot junctions},\ }\href {https://doi.org/10.1088/1361-648X/adbeaf} {\bibfield  {journal} {\bibinfo  {journal} {J. Phys.: Condens. Matter}\ }\textbf {\bibinfo {volume} {37}},\ \bibinfo {pages} {175301} (\bibinfo {year} {2025})}\BibitemShut {NoStop}%
\bibitem [{\citenamefont {Zhao}\ \emph {et~al.}(2023)\citenamefont {Zhao}, \citenamefont {Cui}, \citenamefont {Volkov}, \citenamefont {Yoo}, \citenamefont {Lee}, \citenamefont {Gardener}, \citenamefont {Akey}, \citenamefont {Engelke}, \citenamefont {Ronen}, \citenamefont {Zhong}, \citenamefont {Gu}, \citenamefont {Plugge}, \citenamefont {Tummuru}, \citenamefont {Kim}, \citenamefont {Franz}, \citenamefont {Pixley}, \citenamefont {Poccia},\ and\ \citenamefont {Kim}}]{Zhao2023}%
  \BibitemOpen
  \bibfield  {author} {\bibinfo {author} {\bibfnamefont {S.~Y.~F.}\ \bibnamefont {Zhao}}, \bibinfo {author} {\bibfnamefont {X.}~\bibnamefont {Cui}}, \bibinfo {author} {\bibfnamefont {P.~A.}\ \bibnamefont {Volkov}}, \bibinfo {author} {\bibfnamefont {H.}~\bibnamefont {Yoo}}, \bibinfo {author} {\bibfnamefont {S.}~\bibnamefont {Lee}}, \bibinfo {author} {\bibfnamefont {J.~A.}\ \bibnamefont {Gardener}}, \bibinfo {author} {\bibfnamefont {A.~J.}\ \bibnamefont {Akey}}, \bibinfo {author} {\bibfnamefont {R.}~\bibnamefont {Engelke}}, \bibinfo {author} {\bibfnamefont {Y.}~\bibnamefont {Ronen}}, \bibinfo {author} {\bibfnamefont {R.}~\bibnamefont {Zhong}}, \bibinfo {author} {\bibfnamefont {G.}~\bibnamefont {Gu}}, \bibinfo {author} {\bibfnamefont {S.}~\bibnamefont {Plugge}}, \bibinfo {author} {\bibfnamefont {T.}~\bibnamefont {Tummuru}}, \bibinfo {author} {\bibfnamefont {M.}~\bibnamefont {Kim}}, \bibinfo {author} {\bibfnamefont {M.}~\bibnamefont {Franz}}, \bibinfo {author} {\bibfnamefont {J.~H.}\ \bibnamefont {Pixley}},
  \bibinfo {author} {\bibfnamefont {N.}~\bibnamefont {Poccia}},\ and\ \bibinfo {author} {\bibfnamefont {P.}~\bibnamefont {Kim}},\ }\bibfield  {title} {\bibinfo {title} {Time-reversal symmetry breaking superconductivity between twisted cuprate superconductors},\ }\href {https://doi.org/10.1126/science.abl8371} {\bibfield  {journal} {\bibinfo  {journal} {Science}\ }\textbf {\bibinfo {volume} {382}},\ \bibinfo {pages} {1422} (\bibinfo {year} {2023})},\ \Eprint {https://arxiv.org/abs/https://www.science.org/doi/pdf/10.1126/science.abl8371} {https://www.science.org/doi/pdf/10.1126/science.abl8371} \BibitemShut {NoStop}%
\bibitem [{\citenamefont {Yu}\ \emph {et~al.}(2024)\citenamefont {Yu}, \citenamefont {Cuozzo}, \citenamefont {Sapkota}, \citenamefont {Rossi}, \citenamefont {Rademacher}, \citenamefont {Nenoff},\ and\ \citenamefont {Pan}}]{Yu2024}%
  \BibitemOpen
  \bibfield  {author} {\bibinfo {author} {\bibfnamefont {W.}~\bibnamefont {Yu}}, \bibinfo {author} {\bibfnamefont {J.~J.}\ \bibnamefont {Cuozzo}}, \bibinfo {author} {\bibfnamefont {K.}~\bibnamefont {Sapkota}}, \bibinfo {author} {\bibfnamefont {E.}~\bibnamefont {Rossi}}, \bibinfo {author} {\bibfnamefont {D.~X.}\ \bibnamefont {Rademacher}}, \bibinfo {author} {\bibfnamefont {T.~M.}\ \bibnamefont {Nenoff}},\ and\ \bibinfo {author} {\bibfnamefont {W.}~\bibnamefont {Pan}},\ }\bibfield  {title} {\bibinfo {title} {Time reversal symmetry breaking and zero magnetic field josephson diode effect in dirac semimetal $\mathrm{C}{\mathrm{d}}_{3}\mathrm{A}{\mathrm{s}}_{2}$ mediated asymmetric squids},\ }\href {https://doi.org/10.1103/PhysRevB.110.104510} {\bibfield  {journal} {\bibinfo  {journal} {Phys. Rev. B}\ }\textbf {\bibinfo {volume} {110}},\ \bibinfo {pages} {104510} (\bibinfo {year} {2024})}\BibitemShut {NoStop}%
\bibitem [{\citenamefont {Qiu}\ \emph {et~al.}(2023)\citenamefont {Qiu}, \citenamefont {Yang}, \citenamefont {Hu}, \citenamefont {Zhang}, \citenamefont {Chen}, \citenamefont {Lyu}, \citenamefont {Eckberg}, \citenamefont {Deng}, \citenamefont {Krylyuk}, \citenamefont {Davydov}, \citenamefont {Zhang},\ and\ \citenamefont {Wang}}]{qiu_emergent_2023}%
  \BibitemOpen
  \bibfield  {author} {\bibinfo {author} {\bibfnamefont {G.}~\bibnamefont {Qiu}}, \bibinfo {author} {\bibfnamefont {H.-Y.}\ \bibnamefont {Yang}}, \bibinfo {author} {\bibfnamefont {L.}~\bibnamefont {Hu}}, \bibinfo {author} {\bibfnamefont {H.}~\bibnamefont {Zhang}}, \bibinfo {author} {\bibfnamefont {C.-Y.}\ \bibnamefont {Chen}}, \bibinfo {author} {\bibfnamefont {Y.}~\bibnamefont {Lyu}}, \bibinfo {author} {\bibfnamefont {C.}~\bibnamefont {Eckberg}}, \bibinfo {author} {\bibfnamefont {P.}~\bibnamefont {Deng}}, \bibinfo {author} {\bibfnamefont {S.}~\bibnamefont {Krylyuk}}, \bibinfo {author} {\bibfnamefont {A.~V.}\ \bibnamefont {Davydov}}, \bibinfo {author} {\bibfnamefont {R.}~\bibnamefont {Zhang}},\ and\ \bibinfo {author} {\bibfnamefont {K.~L.}\ \bibnamefont {Wang}},\ }\bibfield  {title} {\bibinfo {title} {Emergent ferromagnetism with superconductivity in fe(te,se) van der waals josephson junctions},\ }\href {https://doi.org/10.1038/s41467-023-42447-4} {\bibfield  {journal} {\bibinfo  {journal} {Nature
  Communications}\ }\textbf {\bibinfo {volume} {14}},\ \bibinfo {pages} {6691} (\bibinfo {year} {2023})}\BibitemShut {NoStop}%
\bibitem [{\citenamefont {Díez-Mérida}\ \emph {et~al.}(2023)\citenamefont {Díez-Mérida}, \citenamefont {Díez-Carlón}, \citenamefont {Yang}, \citenamefont {Xie}, \citenamefont {Gao}, \citenamefont {Senior}, \citenamefont {Watanabe}, \citenamefont {Taniguchi}, \citenamefont {Lu}, \citenamefont {Higginbotham}, \citenamefont {Law},\ and\ \citenamefont {Efetov}}]{diez-merida_symmetry-broken_2023}%
  \BibitemOpen
  \bibfield  {author} {\bibinfo {author} {\bibfnamefont {J.}~\bibnamefont {Díez-Mérida}}, \bibinfo {author} {\bibfnamefont {A.}~\bibnamefont {Díez-Carlón}}, \bibinfo {author} {\bibfnamefont {S.~Y.}\ \bibnamefont {Yang}}, \bibinfo {author} {\bibfnamefont {Y.-M.}\ \bibnamefont {Xie}}, \bibinfo {author} {\bibfnamefont {X.-J.}\ \bibnamefont {Gao}}, \bibinfo {author} {\bibfnamefont {J.}~\bibnamefont {Senior}}, \bibinfo {author} {\bibfnamefont {K.}~\bibnamefont {Watanabe}}, \bibinfo {author} {\bibfnamefont {T.}~\bibnamefont {Taniguchi}}, \bibinfo {author} {\bibfnamefont {X.}~\bibnamefont {Lu}}, \bibinfo {author} {\bibfnamefont {A.~P.}\ \bibnamefont {Higginbotham}}, \bibinfo {author} {\bibfnamefont {K.~T.}\ \bibnamefont {Law}},\ and\ \bibinfo {author} {\bibfnamefont {D.~K.}\ \bibnamefont {Efetov}},\ }\bibfield  {title} {\bibinfo {title} {Symmetry-broken {Josephson} junctions and superconducting diodes in magic-angle twisted bilayer graphene},\ }\href {https://doi.org/10.1038/s41467-023-38005-7} {\bibfield
  {journal} {\bibinfo  {journal} {Nature Communications}\ }\textbf {\bibinfo {volume} {14}},\ \bibinfo {pages} {2396} (\bibinfo {year} {2023})},\ \bibinfo {note} {publisher: Nature Publishing Group}\BibitemShut {NoStop}%
\bibitem [{\citenamefont {Lin}\ \emph {et~al.}(2022)\citenamefont {Lin}, \citenamefont {Siriviboon}, \citenamefont {Scammell}, \citenamefont {Liu}, \citenamefont {Rhodes}, \citenamefont {Watanabe}, \citenamefont {Taniguchi}, \citenamefont {Hone}, \citenamefont {Scheurer},\ and\ \citenamefont {Li}}]{lin_zero-field_2022}%
  \BibitemOpen
  \bibfield  {author} {\bibinfo {author} {\bibfnamefont {J.-X.}\ \bibnamefont {Lin}}, \bibinfo {author} {\bibfnamefont {P.}~\bibnamefont {Siriviboon}}, \bibinfo {author} {\bibfnamefont {H.~D.}\ \bibnamefont {Scammell}}, \bibinfo {author} {\bibfnamefont {S.}~\bibnamefont {Liu}}, \bibinfo {author} {\bibfnamefont {D.}~\bibnamefont {Rhodes}}, \bibinfo {author} {\bibfnamefont {K.}~\bibnamefont {Watanabe}}, \bibinfo {author} {\bibfnamefont {T.}~\bibnamefont {Taniguchi}}, \bibinfo {author} {\bibfnamefont {J.}~\bibnamefont {Hone}}, \bibinfo {author} {\bibfnamefont {M.~S.}\ \bibnamefont {Scheurer}},\ and\ \bibinfo {author} {\bibfnamefont {J.}~\bibnamefont {Li}},\ }\bibfield  {title} {\bibinfo {title} {Zero-field superconducting diode effect in small-twist-angle trilayer graphene},\ }\href {https://doi.org/10.1038/s41567-022-01700-1} {\bibfield  {journal} {\bibinfo  {journal} {Nature Physics}\ }\textbf {\bibinfo {volume} {18}},\ \bibinfo {pages} {1221} (\bibinfo {year} {2022})},\ \bibinfo {note} {publisher: Nature
  Publishing Group}\BibitemShut {NoStop}%
\bibitem [{\citenamefont {Wu}\ \emph {et~al.}(2022{\natexlab{b}})\citenamefont {Wu}, \citenamefont {Wang}, \citenamefont {Xu}, \citenamefont {Sivakumar}, \citenamefont {Pasco}, \citenamefont {Filippozzi}, \citenamefont {Parkin}, \citenamefont {Zeng}, \citenamefont {McQueen},\ and\ \citenamefont {Ali}}]{wu_field-free_2022}%
  \BibitemOpen
  \bibfield  {author} {\bibinfo {author} {\bibfnamefont {H.}~\bibnamefont {Wu}}, \bibinfo {author} {\bibfnamefont {Y.}~\bibnamefont {Wang}}, \bibinfo {author} {\bibfnamefont {Y.}~\bibnamefont {Xu}}, \bibinfo {author} {\bibfnamefont {P.~K.}\ \bibnamefont {Sivakumar}}, \bibinfo {author} {\bibfnamefont {C.}~\bibnamefont {Pasco}}, \bibinfo {author} {\bibfnamefont {U.}~\bibnamefont {Filippozzi}}, \bibinfo {author} {\bibfnamefont {S.~S.~P.}\ \bibnamefont {Parkin}}, \bibinfo {author} {\bibfnamefont {Y.-J.}\ \bibnamefont {Zeng}}, \bibinfo {author} {\bibfnamefont {T.}~\bibnamefont {McQueen}},\ and\ \bibinfo {author} {\bibfnamefont {M.~N.}\ \bibnamefont {Ali}},\ }\bibfield  {title} {\bibinfo {title} {The field-free {Josephson} diode in a van der {Waals} heterostructure},\ }\href {https://doi.org/10.1038/s41586-022-04504-8} {\bibfield  {journal} {\bibinfo  {journal} {Nature}\ }\textbf {\bibinfo {volume} {604}},\ \bibinfo {pages} {653} (\bibinfo {year} {2022}{\natexlab{b}})},\ \bibinfo {note} {publisher: Nature
  Publishing Group}\BibitemShut {NoStop}%
\bibitem [{\citenamefont {Liu}\ \emph {et~al.}(2024)\citenamefont {Liu}, \citenamefont {Itahashi}, \citenamefont {Aoki}, \citenamefont {Dong}, \citenamefont {Wang}, \citenamefont {Ogawa}, \citenamefont {Ideue},\ and\ \citenamefont {Iwasa}}]{Liu2024_strained}%
  \BibitemOpen
  \bibfield  {author} {\bibinfo {author} {\bibfnamefont {F.}~\bibnamefont {Liu}}, \bibinfo {author} {\bibfnamefont {Y.~M.}\ \bibnamefont {Itahashi}}, \bibinfo {author} {\bibfnamefont {S.}~\bibnamefont {Aoki}}, \bibinfo {author} {\bibfnamefont {Y.}~\bibnamefont {Dong}}, \bibinfo {author} {\bibfnamefont {Z.}~\bibnamefont {Wang}}, \bibinfo {author} {\bibfnamefont {N.}~\bibnamefont {Ogawa}}, \bibinfo {author} {\bibfnamefont {T.}~\bibnamefont {Ideue}},\ and\ \bibinfo {author} {\bibfnamefont {Y.}~\bibnamefont {Iwasa}},\ }\bibfield  {title} {\bibinfo {title} {Superconducting diode effect under time-reversal symmetry},\ }\href {https://doi.org/10.1126/sciadv.ado1502} {\bibfield  {journal} {\bibinfo  {journal} {Science Advances}\ }\textbf {\bibinfo {volume} {10}},\ \bibinfo {pages} {eado1502} (\bibinfo {year} {2024})},\ \bibinfo {note} {doi: 10.1126/sciadv.ado1502}\BibitemShut {NoStop}%
\bibitem [{\citenamefont {Yang}\ \emph {et~al.}(2024)\citenamefont {Yang}, \citenamefont {Cuozzo}, \citenamefont {Bokka}, \citenamefont {Qiu}, \citenamefont {Eckberg}, \citenamefont {Lyu}, \citenamefont {Huyan}, \citenamefont {Chu}, \citenamefont {Watanabe}, \citenamefont {Taniguchi},\ and\ \citenamefont {Wang}}]{Yang2025}%
  \BibitemOpen
  \bibfield  {author} {\bibinfo {author} {\bibfnamefont {H.-Y.}\ \bibnamefont {Yang}}, \bibinfo {author} {\bibfnamefont {J.~J.}\ \bibnamefont {Cuozzo}}, \bibinfo {author} {\bibfnamefont {A.~J.}\ \bibnamefont {Bokka}}, \bibinfo {author} {\bibfnamefont {G.}~\bibnamefont {Qiu}}, \bibinfo {author} {\bibfnamefont {C.}~\bibnamefont {Eckberg}}, \bibinfo {author} {\bibfnamefont {Y.}~\bibnamefont {Lyu}}, \bibinfo {author} {\bibfnamefont {S.}~\bibnamefont {Huyan}}, \bibinfo {author} {\bibfnamefont {C.-W.}\ \bibnamefont {Chu}}, \bibinfo {author} {\bibfnamefont {K.}~\bibnamefont {Watanabe}}, \bibinfo {author} {\bibfnamefont {T.}~\bibnamefont {Taniguchi}},\ and\ \bibinfo {author} {\bibfnamefont {K.~L.}\ \bibnamefont {Wang}},\ }\href@noop {} {\bibinfo {title} {Field-resilient supercurrent diode in a multiferroic josephson junction}} (\bibinfo {year} {2024}),\ \Eprint {https://arxiv.org/abs/2412.12344} {arXiv:2412.12344 [cond-mat.mes-hall]} \BibitemShut {NoStop}%
\bibitem [{\citenamefont {Onsager}(1931)}]{Onsager1931}%
  \BibitemOpen
  \bibfield  {author} {\bibinfo {author} {\bibfnamefont {L.}~\bibnamefont {Onsager}},\ }\bibfield  {title} {\bibinfo {title} {Reciprocal relations in irreversible processes. i.},\ }\href {https://doi.org/10.1103/PhysRev.37.405} {\bibfield  {journal} {\bibinfo  {journal} {Phys. Rev.}\ }\textbf {\bibinfo {volume} {37}},\ \bibinfo {pages} {405} (\bibinfo {year} {1931})}\BibitemShut {NoStop}%
\bibitem [{\citenamefont {Kubo}(1957)}]{Kubo1957}%
  \BibitemOpen
  \bibfield  {author} {\bibinfo {author} {\bibfnamefont {R.}~\bibnamefont {Kubo}},\ }\bibfield  {title} {\bibinfo {title} {Statistical-mechanical theory of irreversible processes. i. general theory and simple applications to magnetic and conduction problems},\ }\href {https://doi.org/10.1143/JPSJ.12.570} {\bibfield  {journal} {\bibinfo  {journal} {Journal of the Physical Society of Japan}\ }\textbf {\bibinfo {volume} {12}},\ \bibinfo {pages} {570} (\bibinfo {year} {1957})},\ \Eprint {https://arxiv.org/abs/https://doi.org/10.1143/JPSJ.12.570} {https://doi.org/10.1143/JPSJ.12.570} \BibitemShut {NoStop}%
\bibitem [{\citenamefont {Rikken}\ \emph {et~al.}(2001)\citenamefont {Rikken}, \citenamefont {F\"olling},\ and\ \citenamefont {Wyder}}]{Rikken2001}%
  \BibitemOpen
  \bibfield  {author} {\bibinfo {author} {\bibfnamefont {G.~L. J.~A.}\ \bibnamefont {Rikken}}, \bibinfo {author} {\bibfnamefont {J.}~\bibnamefont {F\"olling}},\ and\ \bibinfo {author} {\bibfnamefont {P.}~\bibnamefont {Wyder}},\ }\bibfield  {title} {\bibinfo {title} {Electrical magnetochiral anisotropy},\ }\href {https://doi.org/10.1103/PhysRevLett.87.236602} {\bibfield  {journal} {\bibinfo  {journal} {Phys. Rev. Lett.}\ }\textbf {\bibinfo {volume} {87}},\ \bibinfo {pages} {236602} (\bibinfo {year} {2001})}\BibitemShut {NoStop}%
\bibitem [{\citenamefont {Cuozzo}\ \emph {et~al.}(2024)\citenamefont {Cuozzo}, \citenamefont {Pan}, \citenamefont {Shabani},\ and\ \citenamefont {Rossi}}]{Cuozzo2024_squid}%
  \BibitemOpen
  \bibfield  {author} {\bibinfo {author} {\bibfnamefont {J.~J.}\ \bibnamefont {Cuozzo}}, \bibinfo {author} {\bibfnamefont {W.}~\bibnamefont {Pan}}, \bibinfo {author} {\bibfnamefont {J.}~\bibnamefont {Shabani}},\ and\ \bibinfo {author} {\bibfnamefont {E.}~\bibnamefont {Rossi}},\ }\bibfield  {title} {\bibinfo {title} {Microwave-tunable diode effect in asymmetric squids with topological josephson junctions},\ }\href {https://doi.org/10.1103/PhysRevResearch.6.023011} {\bibfield  {journal} {\bibinfo  {journal} {Phys. Rev. Res.}\ }\textbf {\bibinfo {volume} {6}},\ \bibinfo {pages} {023011} (\bibinfo {year} {2024})}\BibitemShut {NoStop}%
\bibitem [{\citenamefont {Valentini}\ \emph {et~al.}(2024)\citenamefont {Valentini}, \citenamefont {Sagi}, \citenamefont {Baghumyan}, \citenamefont {de~Gijsel}, \citenamefont {Jung}, \citenamefont {Calcaterra}, \citenamefont {Ballabio}, \citenamefont {Aguilera~Servin}, \citenamefont {Aggarwal}, \citenamefont {Janik}, \citenamefont {Adletzberger}, \citenamefont {Seoane~Souto}, \citenamefont {Leijnse}, \citenamefont {Danon}, \citenamefont {Schrade}, \citenamefont {Bakkers}, \citenamefont {Chrastina}, \citenamefont {Isella},\ and\ \citenamefont {Katsaros}}]{Valentini2024}%
  \BibitemOpen
  \bibfield  {author} {\bibinfo {author} {\bibfnamefont {M.}~\bibnamefont {Valentini}}, \bibinfo {author} {\bibfnamefont {O.}~\bibnamefont {Sagi}}, \bibinfo {author} {\bibfnamefont {L.}~\bibnamefont {Baghumyan}}, \bibinfo {author} {\bibfnamefont {T.}~\bibnamefont {de~Gijsel}}, \bibinfo {author} {\bibfnamefont {J.}~\bibnamefont {Jung}}, \bibinfo {author} {\bibfnamefont {S.}~\bibnamefont {Calcaterra}}, \bibinfo {author} {\bibfnamefont {A.}~\bibnamefont {Ballabio}}, \bibinfo {author} {\bibfnamefont {J.}~\bibnamefont {Aguilera~Servin}}, \bibinfo {author} {\bibfnamefont {K.}~\bibnamefont {Aggarwal}}, \bibinfo {author} {\bibfnamefont {M.}~\bibnamefont {Janik}}, \bibinfo {author} {\bibfnamefont {T.}~\bibnamefont {Adletzberger}}, \bibinfo {author} {\bibfnamefont {R.}~\bibnamefont {Seoane~Souto}}, \bibinfo {author} {\bibfnamefont {M.}~\bibnamefont {Leijnse}}, \bibinfo {author} {\bibfnamefont {J.}~\bibnamefont {Danon}}, \bibinfo {author} {\bibfnamefont {C.}~\bibnamefont {Schrade}}, \bibinfo {author} {\bibfnamefont
  {E.}~\bibnamefont {Bakkers}}, \bibinfo {author} {\bibfnamefont {D.}~\bibnamefont {Chrastina}}, \bibinfo {author} {\bibfnamefont {G.}~\bibnamefont {Isella}},\ and\ \bibinfo {author} {\bibfnamefont {G.}~\bibnamefont {Katsaros}},\ }\bibfield  {title} {\bibinfo {title} {Parity-conserving cooper-pair transport and ideal superconducting diode in planar germanium},\ }\href {https://doi.org/10.1038/s41467-023-44114-0} {\bibfield  {journal} {\bibinfo  {journal} {Nature Communications}\ }\textbf {\bibinfo {volume} {15}},\ \bibinfo {pages} {169} (\bibinfo {year} {2024})}\BibitemShut {NoStop}%
\bibitem [{\citenamefont {Seoane~Souto}\ \emph {et~al.}(2024)\citenamefont {Seoane~Souto}, \citenamefont {Leijnse}, \citenamefont {Schrade}, \citenamefont {Valentini}, \citenamefont {Katsaros},\ and\ \citenamefont {Danon}}]{Souto2024}%
  \BibitemOpen
  \bibfield  {author} {\bibinfo {author} {\bibfnamefont {R.}~\bibnamefont {Seoane~Souto}}, \bibinfo {author} {\bibfnamefont {M.}~\bibnamefont {Leijnse}}, \bibinfo {author} {\bibfnamefont {C.}~\bibnamefont {Schrade}}, \bibinfo {author} {\bibfnamefont {M.}~\bibnamefont {Valentini}}, \bibinfo {author} {\bibfnamefont {G.}~\bibnamefont {Katsaros}},\ and\ \bibinfo {author} {\bibfnamefont {J.}~\bibnamefont {Danon}},\ }\bibfield  {title} {\bibinfo {title} {Tuning the josephson diode response with an ac current},\ }\href {https://doi.org/10.1103/PhysRevResearch.6.L022002} {\bibfield  {journal} {\bibinfo  {journal} {Phys. Rev. Res.}\ }\textbf {\bibinfo {volume} {6}},\ \bibinfo {pages} {L022002} (\bibinfo {year} {2024})}\BibitemShut {NoStop}%
\bibitem [{\citenamefont {Reinhardt}\ \emph {et~al.}(2024)\citenamefont {Reinhardt}, \citenamefont {Ascherl}, \citenamefont {Costa}, \citenamefont {Berger}, \citenamefont {Gronin}, \citenamefont {Gardner}, \citenamefont {Lindemann}, \citenamefont {Manfra}, \citenamefont {Fabian}, \citenamefont {Kochan}, \citenamefont {Strunk},\ and\ \citenamefont {Paradiso}}]{Reinhardt2024}%
  \BibitemOpen
  \bibfield  {author} {\bibinfo {author} {\bibfnamefont {S.}~\bibnamefont {Reinhardt}}, \bibinfo {author} {\bibfnamefont {T.}~\bibnamefont {Ascherl}}, \bibinfo {author} {\bibfnamefont {A.}~\bibnamefont {Costa}}, \bibinfo {author} {\bibfnamefont {J.}~\bibnamefont {Berger}}, \bibinfo {author} {\bibfnamefont {S.}~\bibnamefont {Gronin}}, \bibinfo {author} {\bibfnamefont {G.~C.}\ \bibnamefont {Gardner}}, \bibinfo {author} {\bibfnamefont {T.}~\bibnamefont {Lindemann}}, \bibinfo {author} {\bibfnamefont {M.~J.}\ \bibnamefont {Manfra}}, \bibinfo {author} {\bibfnamefont {J.}~\bibnamefont {Fabian}}, \bibinfo {author} {\bibfnamefont {D.}~\bibnamefont {Kochan}}, \bibinfo {author} {\bibfnamefont {C.}~\bibnamefont {Strunk}},\ and\ \bibinfo {author} {\bibfnamefont {N.}~\bibnamefont {Paradiso}},\ }\bibfield  {title} {\bibinfo {title} {Link between supercurrent diode and anomalous josephson effect revealed by gate-controlled interferometry},\ }\href {https://doi.org/10.1038/s41467-024-48741-z} {\bibfield  {journal} {\bibinfo
  {journal} {Nature Communications}\ }\textbf {\bibinfo {volume} {15}},\ \bibinfo {pages} {4413} (\bibinfo {year} {2024})}\BibitemShut {NoStop}%
\bibitem [{\citenamefont {Buzdin}(2008)}]{Buzdin2008}%
  \BibitemOpen
  \bibfield  {author} {\bibinfo {author} {\bibfnamefont {A.}~\bibnamefont {Buzdin}},\ }\bibfield  {title} {\bibinfo {title} {Direct coupling between magnetism and superconducting current in the josephson ${\ensuremath{\varphi}}_{0}$ junction},\ }\href {https://doi.org/10.1103/PhysRevLett.101.107005} {\bibfield  {journal} {\bibinfo  {journal} {Phys. Rev. Lett.}\ }\textbf {\bibinfo {volume} {101}},\ \bibinfo {pages} {107005} (\bibinfo {year} {2008})}\BibitemShut {NoStop}%
\bibitem [{\citenamefont {Likharev}(1979)}]{Likharev1979}%
  \BibitemOpen
  \bibfield  {author} {\bibinfo {author} {\bibfnamefont {K.~K.}\ \bibnamefont {Likharev}},\ }\bibfield  {title} {\bibinfo {title} {Superconducting weak links},\ }\href {https://doi.org/10.1103/RevModPhys.51.101} {\bibfield  {journal} {\bibinfo  {journal} {Rev. Mod. Phys.}\ }\textbf {\bibinfo {volume} {51}},\ \bibinfo {pages} {101} (\bibinfo {year} {1979})}\BibitemShut {NoStop}%
\bibitem [{\citenamefont {Kochan}\ \emph {et~al.}(2023)\citenamefont {Kochan}, \citenamefont {Costa}, \citenamefont {Zhumagulov},\ and\ \citenamefont {Zutic}}]{Kochan2023}%
  \BibitemOpen
  \bibfield  {author} {\bibinfo {author} {\bibfnamefont {D.}~\bibnamefont {Kochan}}, \bibinfo {author} {\bibfnamefont {A.}~\bibnamefont {Costa}}, \bibinfo {author} {\bibfnamefont {I.}~\bibnamefont {Zhumagulov}},\ and\ \bibinfo {author} {\bibfnamefont {I.}~\bibnamefont {Zutic}},\ }\href@noop {} {\bibinfo {title} {Phenomenological theory of the supercurrent diode effect: The lifshitz invariant}} (\bibinfo {year} {2023}),\ \Eprint {https://arxiv.org/abs/2303.11975} {arXiv:2303.11975 [cond-mat.supr-con]} \BibitemShut {NoStop}%
\bibitem [{\citenamefont {Mayer}\ \emph {et~al.}(2020)\citenamefont {Mayer}, \citenamefont {Dartiailh}, \citenamefont {Yuan}, \citenamefont {Wickramasinghe}, \citenamefont {Rossi},\ and\ \citenamefont {Shabani}}]{Mayer2020}%
  \BibitemOpen
  \bibfield  {author} {\bibinfo {author} {\bibfnamefont {W.}~\bibnamefont {Mayer}}, \bibinfo {author} {\bibfnamefont {M.~C.}\ \bibnamefont {Dartiailh}}, \bibinfo {author} {\bibfnamefont {J.}~\bibnamefont {Yuan}}, \bibinfo {author} {\bibfnamefont {K.~S.}\ \bibnamefont {Wickramasinghe}}, \bibinfo {author} {\bibfnamefont {E.}~\bibnamefont {Rossi}},\ and\ \bibinfo {author} {\bibfnamefont {J.}~\bibnamefont {Shabani}},\ }\bibfield  {title} {\bibinfo {title} {{Gate controlled anomalous phase shift in Al/InAs Josephson junctions}},\ }\bibfield  {journal} {\bibinfo  {journal} {Nature Communications}\ }\href {https://doi.org/10.1038/s41467-019-14094-1} {10.1038/s41467-019-14094-1} (\bibinfo {year} {2020}),\ \Eprint {https://arxiv.org/abs/1905.12670} {arXiv:1905.12670} \BibitemShut {NoStop}%
\bibitem [{\citenamefont {Banerjee}\ \emph {et~al.}(2023)\citenamefont {Banerjee}, \citenamefont {Geier}, \citenamefont {Rahman}, \citenamefont {Thomas}, \citenamefont {Wang}, \citenamefont {Manfra}, \citenamefont {Flensberg},\ and\ \citenamefont {Marcus}}]{Banerjee2023}%
  \BibitemOpen
  \bibfield  {author} {\bibinfo {author} {\bibfnamefont {A.}~\bibnamefont {Banerjee}}, \bibinfo {author} {\bibfnamefont {M.}~\bibnamefont {Geier}}, \bibinfo {author} {\bibfnamefont {M.~A.}\ \bibnamefont {Rahman}}, \bibinfo {author} {\bibfnamefont {C.}~\bibnamefont {Thomas}}, \bibinfo {author} {\bibfnamefont {T.}~\bibnamefont {Wang}}, \bibinfo {author} {\bibfnamefont {M.~J.}\ \bibnamefont {Manfra}}, \bibinfo {author} {\bibfnamefont {K.}~\bibnamefont {Flensberg}},\ and\ \bibinfo {author} {\bibfnamefont {C.~M.}\ \bibnamefont {Marcus}},\ }\bibfield  {title} {\bibinfo {title} {Phase asymmetry of andreev spectra from cooper-pair momentum},\ }\href {https://doi.org/10.1103/PhysRevLett.131.196301} {\bibfield  {journal} {\bibinfo  {journal} {Phys. Rev. Lett.}\ }\textbf {\bibinfo {volume} {131}},\ \bibinfo {pages} {196301} (\bibinfo {year} {2023})}\BibitemShut {NoStop}%
\bibitem [{\citenamefont {Yokoyama}\ \emph {et~al.}(2014)\citenamefont {Yokoyama}, \citenamefont {Eto},\ and\ \citenamefont {Nazarov}}]{Yokoyama2014}%
  \BibitemOpen
  \bibfield  {author} {\bibinfo {author} {\bibfnamefont {T.}~\bibnamefont {Yokoyama}}, \bibinfo {author} {\bibfnamefont {M.}~\bibnamefont {Eto}},\ and\ \bibinfo {author} {\bibfnamefont {Y.~V.}\ \bibnamefont {Nazarov}},\ }\bibfield  {title} {\bibinfo {title} {Anomalous josephson effect induced by spin-orbit interaction and zeeman effect in semiconductor nanowires},\ }\href {https://doi.org/10.1103/PhysRevB.89.195407} {\bibfield  {journal} {\bibinfo  {journal} {Phys. Rev. B}\ }\textbf {\bibinfo {volume} {89}},\ \bibinfo {pages} {195407} (\bibinfo {year} {2014})}\BibitemShut {NoStop}%
\bibitem [{\citenamefont {Hasan}\ \emph {et~al.}(2024)\citenamefont {Hasan}, \citenamefont {Shaffer}, \citenamefont {Khodas},\ and\ \citenamefont {Levchenko}}]{Hasan2024}%
  \BibitemOpen
  \bibfield  {author} {\bibinfo {author} {\bibfnamefont {J.}~\bibnamefont {Hasan}}, \bibinfo {author} {\bibfnamefont {D.}~\bibnamefont {Shaffer}}, \bibinfo {author} {\bibfnamefont {M.}~\bibnamefont {Khodas}},\ and\ \bibinfo {author} {\bibfnamefont {A.}~\bibnamefont {Levchenko}},\ }\bibfield  {title} {\bibinfo {title} {Supercurrent diode effect in helical superconductors},\ }\href {https://doi.org/10.1103/PhysRevB.110.024508} {\bibfield  {journal} {\bibinfo  {journal} {Phys. Rev. B}\ }\textbf {\bibinfo {volume} {110}},\ \bibinfo {pages} {024508} (\bibinfo {year} {2024})}\BibitemShut {NoStop}%
\bibitem [{\citenamefont {Su}\ \emph {et~al.}(2024)\citenamefont {Su}, \citenamefont {Wang}, \citenamefont {Gao}, \citenamefont {Luo}, \citenamefont {Yan}, \citenamefont {Wu}, \citenamefont {Li}, \citenamefont {Shen}, \citenamefont {Lu}, \citenamefont {Pan}, \citenamefont {Zhao}, \citenamefont {Zhang},\ and\ \citenamefont {Xu}}]{Su2024}%
  \BibitemOpen
  \bibfield  {author} {\bibinfo {author} {\bibfnamefont {H.}~\bibnamefont {Su}}, \bibinfo {author} {\bibfnamefont {J.-Y.}\ \bibnamefont {Wang}}, \bibinfo {author} {\bibfnamefont {H.}~\bibnamefont {Gao}}, \bibinfo {author} {\bibfnamefont {Y.}~\bibnamefont {Luo}}, \bibinfo {author} {\bibfnamefont {S.}~\bibnamefont {Yan}}, \bibinfo {author} {\bibfnamefont {X.}~\bibnamefont {Wu}}, \bibinfo {author} {\bibfnamefont {G.}~\bibnamefont {Li}}, \bibinfo {author} {\bibfnamefont {J.}~\bibnamefont {Shen}}, \bibinfo {author} {\bibfnamefont {L.}~\bibnamefont {Lu}}, \bibinfo {author} {\bibfnamefont {D.}~\bibnamefont {Pan}}, \bibinfo {author} {\bibfnamefont {J.}~\bibnamefont {Zhao}}, \bibinfo {author} {\bibfnamefont {P.}~\bibnamefont {Zhang}},\ and\ \bibinfo {author} {\bibfnamefont {H.~Q.}\ \bibnamefont {Xu}},\ }\bibfield  {title} {\bibinfo {title} {Microwave-assisted unidirectional superconductivity in al-inas nanowire-al junctions under magnetic fields},\ }\href {https://doi.org/10.1103/PhysRevLett.133.087001} {\bibfield
  {journal} {\bibinfo  {journal} {Phys. Rev. Lett.}\ }\textbf {\bibinfo {volume} {133}},\ \bibinfo {pages} {087001} (\bibinfo {year} {2024})}\BibitemShut {NoStop}%
\bibitem [{\citenamefont {Amundsen}\ \emph {et~al.}(2024)\citenamefont {Amundsen}, \citenamefont {Linder}, \citenamefont {Robinson}, \citenamefont {\ifmmode \check{Z}\else \v{Z}\fi{}uti\ifmmode~\acute{c}\else \'{c}\fi{}},\ and\ \citenamefont {Banerjee}}]{Amundsen2024}%
  \BibitemOpen
  \bibfield  {author} {\bibinfo {author} {\bibfnamefont {M.}~\bibnamefont {Amundsen}}, \bibinfo {author} {\bibfnamefont {J.}~\bibnamefont {Linder}}, \bibinfo {author} {\bibfnamefont {J.~W.~A.}\ \bibnamefont {Robinson}}, \bibinfo {author} {\bibfnamefont {I.}~\bibnamefont {\ifmmode \check{Z}\else \v{Z}\fi{}uti\ifmmode~\acute{c}\else \'{c}\fi{}}},\ and\ \bibinfo {author} {\bibfnamefont {N.}~\bibnamefont {Banerjee}},\ }\bibfield  {title} {\bibinfo {title} {Colloquium: Spin-orbit effects in superconducting hybrid structures},\ }\href {https://doi.org/10.1103/RevModPhys.96.021003} {\bibfield  {journal} {\bibinfo  {journal} {Rev. Mod. Phys.}\ }\textbf {\bibinfo {volume} {96}},\ \bibinfo {pages} {021003} (\bibinfo {year} {2024})}\BibitemShut {NoStop}%
\bibitem [{\citenamefont {Hosur}\ and\ \citenamefont {Palacios}(2023)}]{Hosur2023}%
  \BibitemOpen
  \bibfield  {author} {\bibinfo {author} {\bibfnamefont {P.}~\bibnamefont {Hosur}}\ and\ \bibinfo {author} {\bibfnamefont {D.}~\bibnamefont {Palacios}},\ }\bibfield  {title} {\bibinfo {title} {Proximity-induced equilibrium supercurrent and perfect superconducting diode effect due to band asymmetry},\ }\href {https://doi.org/10.1103/PhysRevB.108.094513} {\bibfield  {journal} {\bibinfo  {journal} {Phys. Rev. B}\ }\textbf {\bibinfo {volume} {108}},\ \bibinfo {pages} {094513} (\bibinfo {year} {2023})}\BibitemShut {NoStop}%
\bibitem [{\citenamefont {Cubaynes}\ \emph {et~al.}(2020)\citenamefont {Cubaynes}, \citenamefont {Contamin}, \citenamefont {Dartiailh}, \citenamefont {Desjardins}, \citenamefont {Cottet}, \citenamefont {Delbecq},\ and\ \citenamefont {Kontos}}]{Cubaynes2020}%
  \BibitemOpen
  \bibfield  {author} {\bibinfo {author} {\bibfnamefont {T.}~\bibnamefont {Cubaynes}}, \bibinfo {author} {\bibfnamefont {L.~C.}\ \bibnamefont {Contamin}}, \bibinfo {author} {\bibfnamefont {M.~C.}\ \bibnamefont {Dartiailh}}, \bibinfo {author} {\bibfnamefont {M.~M.}\ \bibnamefont {Desjardins}}, \bibinfo {author} {\bibfnamefont {A.}~\bibnamefont {Cottet}}, \bibinfo {author} {\bibfnamefont {M.~R.}\ \bibnamefont {Delbecq}},\ and\ \bibinfo {author} {\bibfnamefont {T.}~\bibnamefont {Kontos}},\ }\bibfield  {title} {\bibinfo {title} {Nanoassembly technique of carbon nanotubes for hybrid circuit-qed},\ }\href {https://doi.org/10.1063/5.0021838} {\bibfield  {journal} {\bibinfo  {journal} {Applied Physics Letters}\ }\textbf {\bibinfo {volume} {117}},\ \bibinfo {pages} {114001} (\bibinfo {year} {2020})}\BibitemShut {NoStop}%
\bibitem [{\citenamefont {B\"{a}uml}\ \emph {et~al.}(2021)\citenamefont {B\"{a}uml}, \citenamefont {Bauriedl}, \citenamefont {Marganska}, \citenamefont {Grifoni}, \citenamefont {Strunk},\ and\ \citenamefont {Paradiso}}]{Bauml2021}%
  \BibitemOpen
  \bibfield  {author} {\bibinfo {author} {\bibfnamefont {C.}~\bibnamefont {B\"{a}uml}}, \bibinfo {author} {\bibfnamefont {L.}~\bibnamefont {Bauriedl}}, \bibinfo {author} {\bibfnamefont {M.}~\bibnamefont {Marganska}}, \bibinfo {author} {\bibfnamefont {M.}~\bibnamefont {Grifoni}}, \bibinfo {author} {\bibfnamefont {C.}~\bibnamefont {Strunk}},\ and\ \bibinfo {author} {\bibfnamefont {N.}~\bibnamefont {Paradiso}},\ }\bibfield  {title} {\bibinfo {title} {Supercurrent and phase slips in a ballistic carbon nanotube bundle embedded into a van der waals heterostructure},\ }\href {https://doi.org/10.1021/acs.nanolett.1c02565} {\bibfield  {journal} {\bibinfo  {journal} {Nano Letters}\ }\textbf {\bibinfo {volume} {21}},\ \bibinfo {pages} {8627} (\bibinfo {year} {2021})},\ \bibinfo {note} {pMID: 34634912},\ \Eprint {https://arxiv.org/abs/https://doi.org/10.102/acs.nanolett.1c02565} {https://doi.org/10.102/acs.nanolett.1c02565} \BibitemShut {NoStop}%
\end{thebibliography}%

\end{document}